 \documentclass[10pt,
aps,
prd,
amsmath,
amssymb,
floats,
floatfix,
twocolumn,
superscriptaddress,
nofootinbib,
noshowpacs]{revtex4-2}

\def\aj{{AJ}}       

\usepackage{color}
\usepackage{graphicx}
\usepackage{xcolor}
\definecolor{bluc}{cmyk}{1,0.9,0,0}
\definecolor{rossoCP3}{cmyk}{0,.88,.77,.40}
\definecolor{rosso}{cmyk}{0,1,1,0.4}
\definecolor{rossos}{cmyk}{0,1,1,0.55}
\definecolor{rossoc}{cmyk}{0,1,1,0.2}
\definecolor{verdes}{cmyk}{0.92,0,0.59,0.4}
\usepackage[normalem]{ulem}
\usepackage{hyperref}
\hypersetup{colorlinks, bookmarksopen, bookmarksnumbered, citecolor=verdes, linkcolor=bluc, pdfstartview=FitH, urlcolor=rossos}
\usepackage[utf8x]{inputenc}
\usepackage[T1]{fontenc}
\usepackage{subdepth}

\usepackage{multirow}
\newcommand\rst{\bgroup\markoverwith{\textcolor{red}{\rule[0.5ex]{1pt}{1.4pt}}}\ULon}

\def\aap{{A\&A}}                % Astronomy and Astrophysics

\begin{document}
%\title{Anomalous behavior for a scalar coupling in the Gauss-Bonnet term}
%\title{Signatures of anomalous behavior for a scalar coupling in the Gauss-Bonnet term}
\title{Compact stars in Einstein-scalar–Gauss–Bonnet gravity: regular and divergent scalar field configurations}

\author{Roberto D. Alba Q.}
%\email{robertodaniel.711026@fisica.uaz.edu.mx}
\author{Javier Chagoya}
\author{Armando A. Roque}
\affiliation{
 Unidad Acad\'emica de F\'isica, Universidad Aut\'onoma de Zacatecas, %Calzada Solidaridad esquina con Paseo a la Bufa S/N 
 C.P. 98060, Zacatecas, M\'exico.
}
\date{\today}

\begin{abstract}
We investigate static, spherically symmetric  solutions in Einstein-scalar-Gauss-Bonnet gravity non-minimally coupled to a massless real scalar field, both in vacuum and in the presence of fermionic matter. Focusing on a specific quadratic scalar–Gauss–Bonnet coupling, we identify two distinct classes of compact objects: one with a regular scalar field at the origin—connected to general relativity in an appropriate limit—and another {one} with a divergent scalar field at the origin but a regular geometry. We analyze both purely scalar and matter-supported (hybrid) configurations, showing that the former can describe a broad class of compact objects, while the latter can reproduce neutron star–like masses even when modeled with a simple polytropic equation of state. Furthermore, we highlight distinctive phenomenological signatures, including the ability of these stars to exceed known compactness limits and their potential to act as gravitational wave super-emitters. We also examined the motion of test particles non-minimally coupled to the scalar field and showed the existence of stable circular orbits within the Schwarzschild's ISCO and static configurations at finite radii for particles with
zero angular momentum.
\end{abstract}

\maketitle

%----------------------------------------------------------------------------------------
%	INTRODUCTION
%----------------------------------------------------------------------------------------
\section{Introduction}
%----------------------------

In this paper, we consider Einstein-{scalar}-Gauss-Bonnet gravity coupled to a massless real scalar field and investigate the existence of compact objects within this subclass of the theory. In particular, we present and analyze the phenomenology of  static, spherically symmetric configurations composed of scalar and fermion-scalar fields, identifying a set of novel theoretical regular objects characterized by a divergent scalar field at the origin.\footnote{After obtaining our initial results, we became aware that a purely scalar version of these solutions was discovered in Ref.~\cite{Kleihaus:2019rbg}, where it was introduced under the name ``particle-like solutions''. A comprehensive study was subsequently carried out in Ref.~\cite{Kleihaus:2020qwo}. However, to complete our analysis, we have chosen to present our own version of these solutions and to complement their results with our phenomenological study.} 

The Gauss–Bonnet term appears in several gravitational theories. It is one of the topological invariants that can be added to the four-dimensional Einstein–Hilbert action~\cite{Hilbert1915} without modifying the field equations~\cite{Lovelock:1971yv} (see Sec. 1.2 of Ref.~\cite{Fernandes:2022zrq} for a detailed discussion). However, it becomes relevant in several scenarios motivated by the search for ultraviolet and infrared modifications of general relativity, such as topological gravity, regularized four-dimensional Einstein–Gauss–Bonnet gravity, higher-dimensional gravity, and scalar–tensor gravity. One of the initial motivations for the latter was the appearance of non-minimal couplings between a dynamical scalar field and the Gauss–Bonnet term in the low-energy effective actions of string theories (e.g., in heterotic string theory~\cite{Zwiebach:1985uq, Gross:1986mw}), such modifications in four-dimensional spacetime have been widely explored and are commonly referred to as Einstein{-scalar}-Gauss-Bonnet gravity.\footnote{ Note that Einstein-{scalar}-Gauss-Bonnet gravity is a subclass of Horndeski gravity, which is itself related to the generalized Galileon theories of gravity.~\cite{Horndeski:1974wa, Deffayet:horndeski, Deffayet:horndeski2, Kobayashi:2019hrl}}

The scalar coupling to the Gauss–Bonnet term introduces nontrivial modifications to the gravitational dynamics, particularly in the strong-field regime, while preserving second-order field equations and avoiding ghost-like instabilities. This theory has received growing attention for its ability to support scalarized black hole and neutron star solutions (see the review Ref.~\cite{Doneva:2022ewd}, and for recent developments, Refs.~\cite{Silva:2017uqg, Doneva:2017bvd, Minamitsuji:2023uyb, Lara:2024rwa, Hyun:2024sfv, Belkhadria:2025lev}), thereby providing explicit counterexamples to classical no-hair theorems~\cite{Capuano:2023yyh, Yazadjiev:2025ezx, Sotiriou:2013qea}. In a cosmological context, it offers viable mechanisms for both inflation and late-time acceleration, potentially alleviating the need for exotic matter components~\cite{Venikoudis:2021oee, Aoki:2020ila, Fronimos:2021ejo, Zanoletti:2023ori, Odintsov:2025kyw, Odintsov:2025wai}. Moreover, the theory predicts deviations from general relativistic gravitational wave signatures~\cite{Cardoso:2016oxy, Nojiri:2023mbo}. As such, Einstein-scalar-Gauss-Bonnet gravity provides a rich phenomenological framework for exploring modifications to gravity across both astrophysical and cosmological scales.

In this article, we focus on spherically symmetric configurations, identifying stationary solutions characterized by scalar fields that are either regular or non-regular at the origin. We present a novel class of solutions, with and without fermionic matter, that are globally regular---in the sense that the curvature scalars remain finite at all radii---and asymptotically flat, but whose scalar field diverges at the origin. A subset of these solutions connects, in a certain limit, to their general relativity counterparts, while others arise directly as a consequence of the modified gravity framework. The main results discussed correspond to a specific choice of coupling between the scalar field and the Gauss-Bonnet term; however, we also briefly explore alternative couplings that exhibit similar behavior. To characterize and identify distinctive features of this new class of solutions, we study some phenomenological implications ---highlighting regions of parameter space with high compactness and demonstrating the possibility of reproducing neutron star-like objects even when using a polytropic equation of state. A forthcoming work will investigate the scalarization process in these objects. For a related study, see Ref.~\cite{Ventagli:2020rnx, Ventagli:2021ubn}.

The remainder of this paper is organized as follows. In Sec.~\ref{Sec.TFramework}, we present the theoretical framework. We begin in subsection~\ref{Sec.EGBg} with a discussion of Einstein–Gauss–Bonnet gravity, followed in subsection~\ref{Sec.EGBS} by the derivation of the field equations governing static, spherically symmetric spacetimes. We also introduce the specific scalar–Gauss–Bonnet coupling adopted in this work and analyze the regularity conditions at the origin. These conditions constrain the regions of parameter space in which different classes of compact objects exist: one characterized by a regular scalar field at the origin, and the other by a divergent scalar field at the origin but with a regular geometry.

In Sec.~\ref{Sec.EGBss}, we present our numerical results. Subsection~\ref{EGBSS} introduces pure Einstein-scalar-Gauss-Bonnet star solutions, which feature a divergent scalar field at the origin but remain globally regular, with all curvature scalars finite across the entire spacetime. In subsection~\ref{EGBHS}, we present a novel class of globally regular solutions supported by both fermionic and scalar matter, with scalar fields that can be either regular or divergent at the origin. We refer to these configurations as hybrid Einstein–scalar-Gauss–Bonnet stars, due to the presence of mixed matter content. In Sec.~\ref{Sec.Phenomenology}, we explore their connection to general relativity and analyze phenomenological aspects such as mass-radius relations, compactness, astrophysical classification, and the trajectories of massive particles coupled to the scalar field. 

Finally, we conclude in Sec.~\ref{Sec.ConcRemark} with a summary of our main findings. Additional material is provided in the appendices.

%%%%%%%%%%%%%%%%%%%%%%%%%%%%
\section{Theoretical Framework}\label{Sec.TFramework}
%%%%%%%%%%%%%%%%%%%%%%%%%%%%
As was previously pointed out, the only contribution of adding the Gauss-Bonnet Lagrangian
\begin{equation} \label{eq:GB_Term}
\mathcal{L}_{GB}= R^2-4R_{\mu\nu}R^{\mu\nu}+R_{\mu\nu\gamma\beta}R^{\mu\nu\gamma\beta}
\end{equation}
in the Einstein-Hilbert action, for a $4D-$spacetime, is a boundary term in the variation of the action, i.e., it reduces to a  surface term. A non-trivial scenario arises when a real scalar function $f(\phi)$ is coupled to the Gauss-Bonnet Lagrangian\footnote{ Another possibility, is the dimensionally reduced Gauss-Bonnet theory, where a non-vanishing contribution appears by appropriately rescaling and taking the $D \to 4$ limit of the coupling constant, as proposed in~\cite{Glavan:2019inb}. However, it was later established that this proposal does not define a consistent four-dimensional theory; see Refs.~\cite{Gurses:2020ofy, Ai:2020peo, Shu:2020cjw, Mahapatra:2020rds, Arrechea:2020evj} for detailed discussions. For commentary on various proposed remedies to the dimensional regularization procedure, see Refs.~\cite{Aoki:2020lig, Fernandes:2020nbq, Hennigar:2020lsl}.}
\begin{align} \label{eq:lag}
S=\int d^{4}x \sqrt{-g}\bigg[\frac{M_{\textrm{Pl}}^2}{2}R-X+\gamma f(\phi)\mathcal{L}_{GB}\bigg] \,,
\end{align}
where $M_{\textrm{Pl}}:=1/\sqrt{8\pi G}$ is the reduced Planck mass (we work in natural units, $\hbar=c=1$). We decided to include the standard canonical kinetic term $X:= g^{\mu\nu}\nabla_{\mu}\phi\nabla_{\nu}\phi/2$ to model the dynamics of the massless real scalar field $\phi$. The dimensionless constant $\gamma$ takes the value zero when the Gauss-Bonnet contribution is not considered, and one otherwise. Finally, notice that the scalar field $\phi$ has mass dimension one and the scalar function $f(\phi)$ will be dimensionless.

%%%%%%%%%%%%%%%%%%%%%%%%%%
\subsection{Einstein-Gauss-Bonnet gravity}\label{Sec.EGBg}
%%%%%%%%%%%%%%%%%%%%%%%%%%
The combination of the Einstein-Hilbert term with a scalar function coupled to the Gauss-Bonnet term (plus a canonical kinetic term) resulted in the action~(\ref{eq:lag}). This action describes a scalar-tensor gravity which introduces modifications to general relativity due to the extra scalar degree of freedom present in this class of models.

Adding the matter Lagrangian $\mathcal{L}_m[g_{\mu\nu},\Psi]$ (where $\Psi$ represents the matter field) to the action~(\ref{eq:lag}) and taking its variation with respect to the metric $g^{\mu\nu}$ results in
\begin{eqnarray}\label{met_eq}
\xi_{\mu\nu} := G_{\mu\nu} + \frac{2\gamma}{M_{\textrm{Pl}}^2}G^{(\rm{GB})}_{\mu\nu}=\frac{1}{M_{\textrm{Pl}}^2}\left(T_{\mu\nu}+T_{\mu\nu}^{(\phi)}\right)\,,
\end{eqnarray}
where $G_{\mu\nu}:=R_{\mu\nu}-g_{\mu\nu}R/2$ is the Einstein tensor, and $G^{(\rm{GB})}_{\mu\nu}$ represents the gravitational modification introduced by the non-minimal coupling of the Gauss-Bonnet term to the scalar field. In analogy with standard notation, we define the tensor $T_{\mu\nu}^{(\phi)}$ to contain the purely scalar contributions arising from the canonical kinetic term%\footnote{\jf{Quitar esta nota} One might be tempted to consider that the tensor $T_{\mu\nu}^{(\phi)}$ does not constitute a modification to general relativity. However, it is important to point out that the scalar field $\phi$ is introduced with a geometric, rather than particle-base, motivation. Our choice to define $T_{\mu\nu}^{(\phi)}$ follows by analogy with the standard treatment of the kinetic scalar field term.}
, while $T_{\mu\nu}$ denotes the stress-energy tensor of the baryonic matter field:
\begin{subequations}
\begin{eqnarray}
G^{(\rm{GB})}_{\mu\nu} &:=& 8\alpha_{\mu\nu}[g_{\mu\nu}, \nabla\nabla\phi]f_{\phi}+8\beta_{\mu\nu}[g_{\mu\nu}, \nabla\phi]f_{\phi\phi},\label{Eq.GBContrib}\\
T_{\mu\nu}^{(\phi)} &:=& \nabla_{\mu}\phi\nabla_{\nu}\phi-g_{\mu\nu}X,\\
T_{\mu\nu}&:=&\frac{-2}{\sqrt{-g}}\frac{\delta \mathcal{L}_m}{\delta g^{\mu\nu}}\,.\label{Tenerg}
\end{eqnarray}
\end{subequations}
Equation~(\ref{Eq.GBContrib}) depends only on the first and second derivatives of the scalar coupling function $f(\phi)$, denoted by $f_{\phi} := df/d\phi$ and $f_{\phi\phi} := d^2f/d\phi^2$, respectively. The tensor functions $\alpha_{\mu\nu}$ and $\beta_{\mu\nu}$, defined in Eqs.~(\ref{Eq.Ap.Covar1})-(\ref{Eq.Ap.Covar3}) of Appendix~\ref{Ap.Compl.Eq}, depend on second and first covariant derivatives of the scalar field, respectively, and on the metric.

Likewise, the variation with respect to $\phi$ leads to:
\begin{equation}\label{field_eq}
	\Box\phi = -\gamma\mathcal{L}_{GB} f_{\phi}.
\end{equation}
Notice that when $f(\phi)\sim \phi^2$, an effective {non-necessarily constant} mass of the scalar field, generated by the Gauss-Bonnet term Eq.~(\ref{eq:GB_Term}), emerges: $m_{\text{eff}}^2:=\gamma \mathcal{L}_{GB}$. One expects that the scalar field can develop a tachyonic instability in regions where $R^2+R_{\mu\nu\gamma\beta}R^{\mu\nu\gamma\beta}>4R_{\mu\nu}R^{\mu\nu}$. We will return to this discussion later.

From the system~(\ref{met_eq},~\ref{field_eq}), it is clear that setting $\gamma = 0$ reduces the theory to the standard massless Einstein–Klein–Gordon system, where the scalar field has a purely matter-like interpretation. On the other hand, Eq.~(\ref{field_eq}) does not admit constant scalar field solutions $\phi = \phi_0$ unless $f_{\phi}(\phi_0) = 0$ (or $\mathcal{L}_{GB} = 0$ throughout the spacetime). In this paper, we impose this constraint {in order to recover, at some scale, a solution given by the Schwarzschild metric and a constant scalar field--as would be the case for the massless Einstein-Klein-Gordon system.}\footnote{A second condition,  $f_{\phi\phi}(\phi_{0}) = 0$, is necessary to ensure that the Gauss-Bonnet contribution Eq.~(\ref{Eq.GBContrib}) is null. However, under certain assumptions, one can guarantee that these coefficients are small.} This requirement  excludes popular scalar functions such as $f(\phi)\sim \exp(\phi)$~\cite{Kanti:1995vq} and $f(\phi)\sim \phi$~\cite{Sotiriou:2014pfa}.

%%%%%%%%%%%%%%%%%%%%%%%%%%%%
\subsection{Einstein-Gauss-Bonnet stars}\label{Sec.EGBS}
%%%%%%%%%%%%%%%%%%%%%%%%%%%%
From Eq.~(\ref{met_eq}), it is evident that for $\gamma \neq 0$, the description of gravity is no longer governed solely by the Einstein equations. However, in the low-energy regime, general relativity must be recovered as an effective theory. In this context, one expects that the Gauss–Bonnet modification may leave observable imprints in the structure and dynamics of astrophysical compact objects.

We derive the modified Einstein field equations under the assumptions of staticity and spherical symmetry for both the spacetime and the scalar field:
\begin{subequations}\label{Eq.StatAnz}
\begin{align}
ds^{2} & =-N(r)^{2}dt^{2}+g(r)^{2}dr^{2}+r^{2}d\theta^{2}+r^{2}\sin^2\theta d\varphi^{2} \,,\label{metric} \\
\phi&=\phi(r).
\end{align}
\end{subequations}
Inserting these ansatzes into Eq.~(\ref{met_eq}), and after the appropriate manipulations, we can write the nontrivial components of the field equations, $\xi_{00}, \xi_{11}$, and $\xi_{22}$ as follows:\footnote{ We assume that the components of the covariant stress-energy tensor ${T^{2}}_{2}={T^{3}}_{3}$, which implies that $\xi_{22}$ and $\xi_{33}$ are linearly dependent. For this reason, we do not include the latter component in Eqs~(\ref{Eq.MotSyst}).}
\begin{widetext}
\begin{subequations}\label{Eq.MotSyst}
\begin{align}
\frac{T_{00}}{N^2 M_{\rm{Pl}}^2}+\frac{\phi'{}^2}{2 M_{\rm{Pl}}^2 g^2}&=\frac{2}{r}\left[1+\frac{4\gamma}{M_{\rm{Pl}}^2}\left(1-\frac{3}{g^2}\right)\frac{f_{\phi}\phi'}{r}\right]\frac{g'}{g^3}-\frac{8\gamma}{M_{\rm{Pl}}^2}\frac{(g^2-1)}{ g^4r^2}\frac{d}{dr}\bigg(f_{\phi}\phi'\bigg)
+\bigg(1-\frac{1}{g^2}\bigg)\frac{1}{r^2}, \label{Eq.MotSyst00}\\[0.2cm]
\frac{T_{11}}{g^2 M_{\rm{Pl}}^2 }+\frac{\phi'{}^2}{2 M_{\rm{Pl}}^2 g^2}&=\left[1+\frac{4\gamma}{M_{\rm{Pl}}^2}\left(1-\frac{3}{g^2}\right)\frac{f_{\phi}\phi'}{r}\right]\frac{N'}{g^2 N}-\left(1-\frac{1}{g^2}\right)\frac{1}{r^2},\label{Eq.MotSyst11}\\[0.2cm]
\frac{g^2 T_{22}}{r^2 M_{\rm{Pl}}^2}-\frac{\phi'{}^2}{2 M_{\rm{Pl}}^2}&=\left(1-\frac{8\gamma}{M_{\rm{Pl}}^2}\frac{f_{\phi}\phi'}{ rg^2}\right)\frac{N''}{N}+\left(\frac{1}{r}-\frac{g'}{g}\right)\left(1-\frac{8\gamma}{M_{\rm{Pl}}^2 g^2}\zeta\left(1-r\frac{g'}{g}\right)^{-1}\right)\frac{N'}{N}-\frac{g'}{g r},
\end{align}
\end{subequations}
\end{widetext}
where the prime denotes derivation with respect to the radial variable. The functions $T_{00}, T_{11}$ and $T_{22}$, are  components of the covariant stress-energy tensor Eq.~(\ref{Tenerg}) and
\begin{align}
    \zeta := \frac{d}{dr}\left(f_{\phi}\phi'\right)-3f_{\phi}\phi'\frac{g'}{g}.
\end{align}

Likewise, from the field equation~(\ref{field_eq}) we arrive to:
\begin{align}\label{eq.KG.modified}
\phi''+\bigg(\frac{N'}{N}-\frac{g'}{g}+\frac{2}{r}\bigg)\phi'+\frac{8\gamma}{r^2}Z f_{\phi}=0,
\end{align}
with $Z$ defined as,
\begin{align}
Z := \frac{N''}{N}\bigg(\frac{1}{g^2}-1\bigg)+\frac{g'}{g}\frac{N'}{N}\bigg(1-\frac{3}{g^2}\bigg).
\end{align}

The field equations~(\ref{Eq.MotSyst},~\ref{eq.KG.modified}) are time-independent, which implies that a rescaling of the lapse function, $N \to \lambda N$, with $\lambda$ an arbitrary constant,  leaves the system invariant. Additionally, one can show that within the system~(\ref{Eq.MotSyst}), only Eqs.~(\ref{Eq.MotSyst00}) and~(\ref{Eq.MotSyst11}) are linearly independent.

In this paper, we consider the baryonic matter field to be composed of fermionic matter. This fermionic component supports no transverse stresses and exhibits no bulk motion, consistent with our aim of studying equilibrium configurations. Under these assumptions, the fermionic energy–momentum tensor~(\ref{Tenerg}) takes the form of a perfect fluid~\cite{Oppenheimer:1939ne, Weinberg:1972kfs, misner2017gravitation}:
\begin{equation}
	T_{\mu\nu}=(\epsilon+p)u_{\mu}u_{\nu}+p g_{\mu\nu}\,,\label{perfl}
\end{equation}
where $p$ is the fluid pressure, and  $\epsilon$ is the total energy density, which includes both the rest mass density $\rho$ and internal energy $\varepsilon$, such that $\epsilon = \rho + \varepsilon$. The assumptions of time independence and spherical symmetry imply that both $p$ and $\epsilon$ are functions of only the radial coordinate. The four-velocity of the fluid is given by $u^{\nu} = u^{0}(1, 0, 0, 0)$, with the component $u^{0}$ determined from the normalization condition, $u^{\nu} u_{\nu} = -1$, which yields $u^{0} = 1/N$.

In line with the goal of this subsection, we need to define the physical conditions that characterize the compact objects studied in this work (hereafter referred to as stars). First, the spacetime curvature must remain finite at $r = 0$. Second, the \textit{total energy} must be finite, ensuring that the Schwarzschild solution is recovered asymptotically. Under our spacetime ansatz, the total energy coincides with the ADM mass (recall that we are using units where $c = 1$)~\cite{PhysRev.122.997}.

%%%%%%%%%%%%%%%%%%%%%%%%%%%%
\subsubsection{The scalar function}
%%%%%%%%%%%%%%%%%%%%%%%%%%%%
Now we proceed to define the scalar function $f(\phi)$, which must satisfy certain constraints. {First of all, we want an $f(\phi)$ that is not everywhere constant since in that case the field equations~(\ref{Eq.MotSyst},~\ref{eq.KG.modified}) reduce to those of general relativity minimally coupled to a canonical scalar field. However, we do impose that there is a region where $\phi$ approaches some asymptotic value $\phi\simeq\phi_0$, such that $f_{\phi}(\phi_0)\simeq0$,  in order to approximately recover the Schwarzschild solution}. Second, {since $\mathcal L_{GB}$ is not positive-definite, Eq.~(\ref{field_eq}) implies that}, a possible tachyonic instability may arise because of the curvature of the spacetime, particularly at higher curvatures. This can be understood by computing the equation corresponding to the linear perturbation $\delta\phi$ on a fixed background,\footnote{ It is necessary to note that Eq.~(\ref{Eq.InestTaq}) is obtained using a linear perturbation method, which restricts its validity to the linear regime, and this may not be the case in regions of higher curvature. However, the equation can still be used to illustrate the emergence of tachyonic instability.}
\begin{align}\label{Eq.InestTaq}
    \left(\Box + \frac{m^{2}_{\text{eff}}}{2}f_{\phi\phi}(\phi_0)\right)\delta\phi=0,
\end{align}
recalling $m_{\text{eff}}^2=\gamma \mathcal{L}_{GB}$. Regions where the condition $m^{2}_{\text{eff}} f_{\phi\phi}(\phi_0) < 0$ holds lead to an imaginary mass, which gives rise to scalar modes that grow exponentially. For example, static neutron stars have regions of both, positive and negative $m^{2}_{\text{eff}}$ (see Fig. 3 in Ref.~\cite{Silva:2017uqg}). The appearance of these tachyonic instabilities does not necessarily lead to an unstable compact object. In some models, developing the instability could induce spontaneous scalarization (see, for a review, Ref.~\cite{Doneva:2022ewd}). Therefore, our choice of the scalar function $f(\phi)$ must ensure that the theory allows objects characterized by high curvature to scalarize, while at low curvature, the system will be described by general relativity solutions. Our choice,
\begin{align}\label{Eq.FuncPhi}
    f(\phi)=\frac{\beta}{2}\phi^2,
\end{align}
satisfies the two constraints: $f_{\phi}(\phi_0)=\beta \phi_0 = 0$ for the trivial solution $\phi_0=0$, and it exhibits spontaneous scalarization~\cite{Silva:2017uqg, Doneva:2017duq}. Notice that, at the linear level, any choice of $f(\phi)$ that satisfies the condition $f_{\phi}(\phi_0)=0$ is equivalent to~(\ref{Eq.FuncPhi}).
The coupling parameter $\beta$ has units of inverse mass squared, and throughout this paper, we restrict our analysis to positive values.  

%%%%%%%%%%%%%%%%%%%%%%%%%%%%
\subsubsection{Finite curvature at the origin}\label{SubSectFinitCurvOrig}
%%%%%%%%%%%%%%%%%%%%%%%%%%%%
The system (\ref{Eq.MotSyst},~\ref{eq.KG.modified}) needs to be complemented by appropriate boundary conditions that describe the physical problem we are addressing: ensuring finite curvature at $r=0$ and the asymptotic recovery of the Schwarzschild solution. Next, we present a discussion about the conditions required for finite curvature at the origin.

From the Ricci scalar associated with the metric ansatz~(\ref{metric}),
\begin{align}\label{Eq.RicciScalar}
R = \left(1-\frac{1}{g^2}\right)\frac{2}{r^2}-\frac{d}{dr}\left(\frac{N}{g}\right)\frac{4}{g N r}-\frac{d}{dr}\left(\frac{N'}{g}\right)\frac{2}{g N},
\end{align}
it is possible to identify, after performing a Taylor expansion around $r = 0$ that, in order to keep the curvature finite, the metric components must satisfy:
\begin{align}\label{Eq.RicciCond}
    g_0=1, \;\quad g_{0}'=\frac{N_0'}{2N_0}\,,
\end{align}
where the subscript $0$ denotes evaluation at $r = 0$, e.g., $g_0':= g'(r)|_{r=0}$. 

Expanding the dynamic equations~(\ref{Eq.MotSyst}) and~(\ref{eq.KG.modified}) in a Taylor series around $r = 0$, and applying the constraints~(\ref{Eq.RicciCond}), one finds (after solving order by order in $r$) that the first derivatives of the metric functions and the scalar field vanish at the origin: $\phi_0'=g_0'(r)=N_0'=0$.\footnote{ A second branch exists for $\phi_0=0$, but this leads to the trivial scalar field solution.} The second derivatives at the origin are then determined by the solutions of the following system:
\begin{subequations}\label{Eq.OrigSecDeriva}
\begin{align}
g_{0}''-\frac{8\gamma f_{\phi_0} g_{0}''\phi_{0}''}{M_{\rm{Pl}}^2}&=\frac{\epsilon}{3 M_{\rm{Pl}}^2},\\
g_{0}''-2\frac{N_{0}''}{N_{0}}+\frac{16\gamma f_{\phi_0} \phi_{0}''}{M_{\rm{Pl}}^2}\frac{N_{0}''}{N_0}&=-\frac{p}{M_{\rm{Pl}}^2},\\
\phi_{0}''-8\gamma f_{\phi_0} g_{0}''\frac{N_{0}''}{N_0}&=0.
\end{align}
\end{subequations}
After some algebraic manipulation, one finds that the system~(\ref{Eq.OrigSecDeriva}) admits four branches of solutions (as functions of $N_0$ and $f_{\phi_0}$) when fermionic matter is present ($\epsilon\neq 0, p\neq 0$).  In the vacuum case ($\epsilon=0, p=0$), the only solution is the trivial one: $g_{0}''=N_{0}''=\phi_{0}''=0$. Proceeding to higher-order terms in the Taylor expansion confirms that $N^{(n)}=g^{(n)}=\phi^{(n)}=0$ for $n=1, 2, \dots$, is the only solution satisfying the full expansion. The asymptotic desired behavior restricts this constant solution to: $N(r)=1, g(r)=1, \phi=\text{arbitrary constant}$. It is important to note that non-trivial solutions exist when fermionic matter is considered, and as shown in Sec.~\ref{EGBHS}, they exhibit the desired asymptotic behavior and correspond to physically meaningful compact objects.

At this point, one might be tempted to conclude that purely scalar-gravity stars are excluded from the model~(\ref{eq:lag}). However, the previous result only indicates that regular radial scalar field profiles at $r = 0$ are not permitted (apart from the trivial constant solution).

To explore the possibility of non-regular but physically acceptable scalar field solutions, we now consider a Laurent series expansion around the origin:
\begin{align}\label{Eq.ScaOrigAnsatz}
    \phi = \sum_{i=-1} a_{i} r^{i}=\frac{a_{-1}}{r}+a_{0}+a_1 r + a_2 r^2 + \mathcal{O}(r^3),
\end{align}
along with the Taylor expansion for the metric components $N(r)$ and $g(r)$. (Note that the ansatz~\eqref{Eq.ScaOrigAnsatz} does not necessarily imply any irregularity in the metric components.) Repeating the previous procedure (imposing the regularity conditions~(\ref{Eq.RicciCond}), expanding the field equations~(\ref{Eq.MotSyst}) and~(\ref{eq.KG.modified}), and solving order by order in $r$), one can identify a class of non-regular scalar field solutions that preserve finite curvature at the origin. The resulting approximate solution, valid up to $\mathcal{O}(r^4)$, is determined in terms of the coefficients $a_{-1}$ and $a_{0}$, with $a_{-1}\neq 0$ (the case $a_{-1}=0$ again leads to the trivial constant solution), and an arbitrary nonzero value of $N_0$,
\begin{widetext}
\begin{subequations}\label{Sys.OrigCond}
\begin{align}
N_{(0)} \approx &\; N_0 + \frac{N_0 r^2}{64\beta} - \frac{N_0 a_0 r^3}{96\beta a_{-1}} +\mathcal{O}(r^4),\\
g_{(0)} \approx &\; 1-\frac{r^2}{32\beta} + \frac{(128\beta a_0^3-27 a_0 a_{-1}^2)r^3}{128\beta(3 a_{-1}^2-64\beta a_0^2)a_{-1}} +\mathcal{O}(r^4),\\
\phi_{(0)} \approx &\; \frac{a_1}{r}+ a_0 + \frac{3 a_{-1} r}{128\beta}+\frac{7a_0a_{-1}^2 r^2}{256\beta(3a_{-1}^2-64\beta a_0^2)} + \frac{a_{-1}(69a_{-1}^2-128\beta a_0^2)r^3}{98304\beta^2(64\beta a_0^2-3a_{-1}^2)} + \mathcal{O}(r^4).
\end{align}
\end{subequations}
\end{widetext}
The finite curvature at $r=0$ is reflected in the behavior of the curvature invariants. For the Ricci ($R$) and Kretschmann ($K$) scalars, we have:
\begin{align}\label{Eq.ScCurvOrig}
    R=-\frac{9}{16\beta \gamma}, \quad K = \frac{15}{(16 \beta\gamma)^2}.
\end{align}
Notably, the values of these curvature invariants depend only on the coupling parameter $\beta$, and not on the coefficients $a_{-1}$ or $a_{0}$. This indicates that the curvature around the origin is determined by the strength of the coupling between the scalar field and the Gauss-Bonnet term, rather than by the specific form of the scalar field profile. Furthermore, in the limit $\beta\to 0$, the invariants~(\ref{Eq.ScCurvOrig}) diverge, suggesting that this branch of solutions is not continuously connected to general relativity. In subsection~\ref{EGBSS}, we explore this class of solutions in more detail.

Now, a natural question arises: can these non-regular scalar field solutions also exist in the presence of fermionic matter, i.e., when $\epsilon\neq 0$ and $p\neq 0$? The answer is yes. By repeating the same procedure used for purely scalar-gravity configurations, we identify a novel class of non-regular scalar field solutions that preserve finite curvature at the origin. In these cases, the curvature invariants at $r=0$ remain given by the expressions in Eq.~(\ref{Eq.ScCurvOrig}).

In summary, we identify two distinct scenarios. The first corresponds to solutions with a regular scalar field at the origin, which exist only when fermionic matter is present (i.e., $\epsilon\neq 0, p\neq 0$). These solutions are determined by the real roots of the system~(\ref{Eq.OrigSecDeriva}), with the scalar function given by Eq.~(\ref{Eq.FuncPhi}) and are continuously connected to general relativity in the limit $\gamma=0$, or equivalently, $\beta=0$. The second scenario features a non-regular scalar field at the origin and arises in both the vacuum ($\epsilon=0, p=0$) and matter-supported cases. These configurations are specific to Einstein-scalar-Gauss-Bonnet gravity and have no analogues in the general relativistic framework.

To conclude, we briefly comment on the relevance of non-regular $\phi(r)$ scalar field profiles. Derrick’s theorem~\cite{10.1063/1.1704233} demonstrates that, in stationary and flat spacetimes, localized solutions to the Klein–Gordon equation are unstable. This result has been extended to the context of general relativity~\cite{Hod:2018dij, Carloni:2019cyo}; however, as shown in Refs.~\cite{Perivolaropoulos:2018cgr, Alestas:2019wtw}, such constraints can be evaded in specific spherically symmetric background geometries. In our work, the theory lies beyond general relativity, rendering these results inapplicable. Nonetheless, they could help explain why the non-regular branch of solutions identified here, is disconnected from the general relativistic limit.

%%%%%%%%%%%%%%%%%%%%%%%%%%%%
\section{Einstein-scalar-Gauss-Bonnet stars}\label{Sec.EGBss}
%%%%%%%%%%%%%%%%%%%%%%%%%%%%

In this section, we present both {pure and hybrid} numerical solutions of the system~(\ref{Eq.MotSyst},~\ref{eq.KG.modified}), with the coupling function $f(\phi)$ given by Eq.~(\ref{Eq.FuncPhi}). We provide several illustrative profiles to highlight the main features of each case. We refer to the vacuum solutions as pure Einstein-scalar-Gauss-Bonnet stars, to distinguish them from the non-vacuum solutions, which we denote as hybrid Einstein-scalar-Gauss-Bonnet stars. To our knowledge, purely scalar solutions were first reported in Refs.~\cite{Kleihaus:2019rbg, Kleihaus:2020qwo}. In this work, we recover those results and extend the discussion to include additional phenomenological aspects. Certain hybrid star configurations involving regular scalar fields---with coupling functions of the form $f(\phi) \sim \exp(-\beta \phi^2) - 1$---have been reported in Refs.~\cite{Doneva:2017duq, Doneva:2023kkz, Kuan:2023trn}, under the name of scalarized neutron stars. These objects are characterized by scalar fields that vanish asymptotically and remain negligible inside the baryonic radius of the star.\footnote{In contrast, we explore compact objects--such as those discussed in Ref.~\cite{Roque:2021lvr}--where the scalar field may reside within or extend beyond the baryonic region.} Black hole solutions within similar scalar–Gauss–Bonnet models have been extensively studied in Refs.~\cite{Doneva:2017bvd, Silva:2017uqg}.

%%%%%%%%%%%%%%%%%%%%%%%%
\subsection{Pure Einstein-scalar-Gauss-Bonnet stars}\label{EGBSS}
%%%%%%%%%%%%%%%%%%%%%%%%
\begin{figure*}
    \centering
    \includegraphics[width=1.\linewidth]{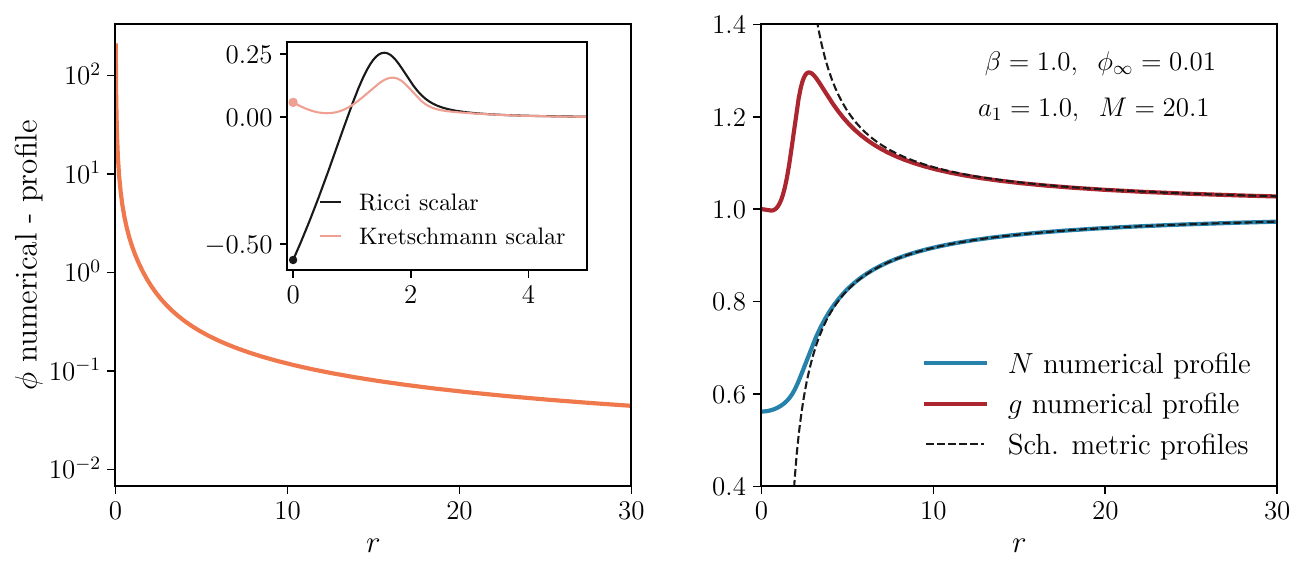}
    \caption{\textbf{Illustrative Einstein-Gauss-Bonnet-scalar stars.} Left panel: radial profile of the scalar field, showing a divergence at the origin, while the Ricci and Kretschmann scalars remain finite (see inset). Circular marker indicates the analytical value from Eqs.~(\ref{Eq.ScCurvOrig}). Right panel: corresponding metric profiles, including Schwarzschild profiles for comparison. The parameters used to compute the boundary conditions at $r = 100$ are specified at the top. See Fig.~\ref{fig:IllExamp2} for details.}\label{fig:IllExamp}
\end{figure*}

\begin{figure*}
    \centering
    \includegraphics[width=1.0\linewidth]{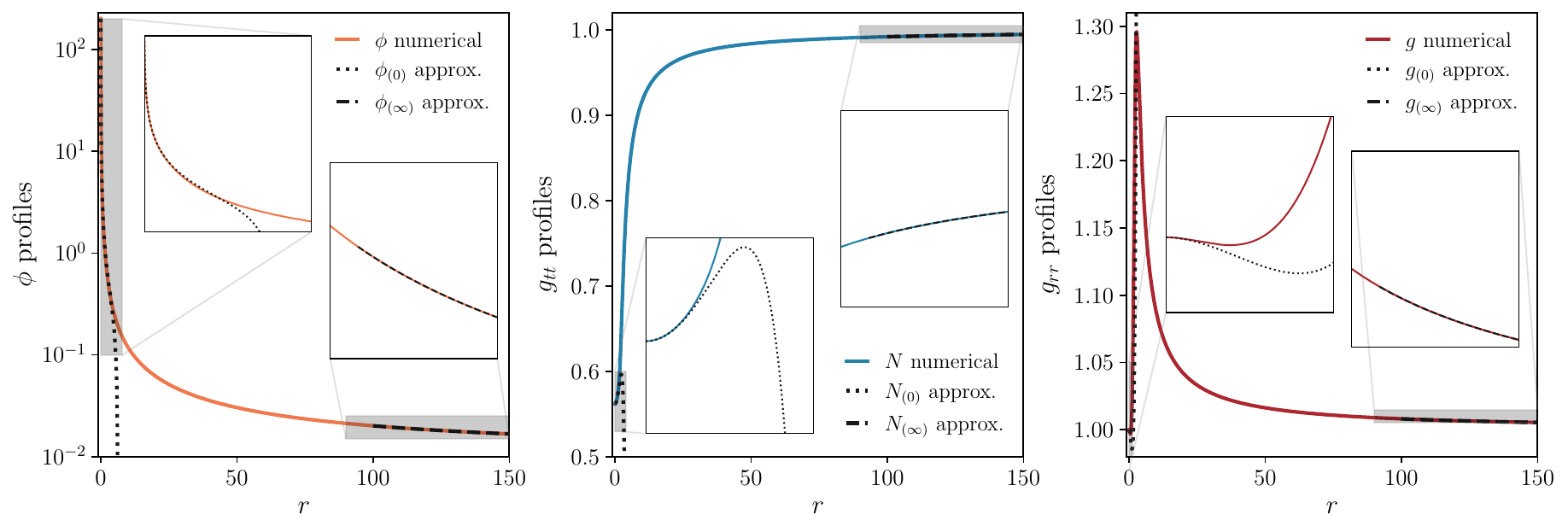}
    \caption{\textbf{Comparison of numerical and approximate solutions.} Full numerical profiles (solid lines) are compared with approximate solutions near the origin (dotted lines, Eqs.~(\ref{Sys.OrigCond})) and at infinity (dashed lines, Eqs.~(\ref{Sys.BoundCond})). These profiles correspond to those shown in Fig.~\ref{fig:IllExamp}. Right panel: notice that although the boundary condition is set at $r = 100$, the numerical solution satisfies $g(r = 0) = 1$, ensuring finite curvature.}\label{fig:IllExamp2}
\end{figure*}

As discussed in the previous section, non-trivial vacuum solutions of the system~(\ref{Eq.MotSyst},~\ref{eq.KG.modified}) require a non-regular scalar field profile at the origin. This prevents establishing boundary conditions at $r = 0$, making it difficult to compute a numerical solution by integrating outward from the origin. One possible approach is to use the regularity conditions~(\ref{Eq.RicciCond}) and integrate outward from a small radius $r_{\text{min}}$, treating the scalar field value $\phi(r_{\text{min}})$ as a shooting parameter, adjusted to satisfy asymptotic flatness at large $r$. An alternative approach (adopted in this work) is to solve the system numerically from a large radius $r_{\text{max}}$ towards the origin. In this case, the boundary conditions are set by evaluating at $r_{\text{max}}$ the asymptotic expansion of the system~(\ref{Eq.MotSyst},~\ref{eq.KG.modified}) 
\begin{widetext}
\begin{subequations}\label{Sys.BoundCond}
\begin{align}
N^2_{(\infty)} \approx &\; 1+\frac{g_1}{r} -\frac{a_1^2 g_1}{12 M_{\rm{Pl}}^2 r^3} + \left(1-\frac{48\beta \phi_{\infty}}{a_1 g_1}\right)\frac{a_1^2 g_1^2}{12  M_{\rm{Pl}}^2 r^4}-\left(1-\frac{a_1^2}{4 g_1^2 M_{\rm{Pl}}^2}-\frac{32\beta \phi_{\infty}}{3 a_1 g_1}\left[1-\frac{4a_1}{a_0g_1}-\frac{a_1^2}{M_{\rm{Pl}}^2 g_1^2}\right]\right)\nonumber\\
    &\times \frac{3a_1^2g_1^3}{40 M_{\rm{Pl}}^2 r^5}+\mathcal{O}(1/r^6),\\
g^{2}_{(\infty)} \approx & \left(1+\frac{g_1}{r}+\frac{a_1^2}{2M_{\rm{Pl}}^2r^2}-\frac{a_1^2 g_1}{4M_{\rm{Pl}}^2r^3}+\left[1-\frac{48\beta \phi_{\infty}}{a_1g_1}\right]\frac{a_1^2g_1^2}{6M_{\rm{Pl}}^2r^4}-\left[1-\frac{a_1^2}{12M_{\rm{Pl}}^2g_1^2}+\frac{64\beta}{g_1^2}\left(1+\frac{\phi_{\infty} a_1}{2M_{\rm{Pl}}^2g_1}\right)\right]\right.\nonumber\\
    & \left. \times \frac{a_1^2g_1^3}{8M_{\rm{Pl}}^2r^5}+\mathcal{O}(1/r^6)\right)^{-1},\\
\phi_{(\infty)} \approx &\; \phi_{\infty} + \frac{a_1}{r} -\frac{a_1 g_1}{2r^2}+\left(1-\frac{a_1^2}{4 M_{\rm{Pl}}^2 g_1^2}\right)\frac{a_1 g_1^2}{3r^3} - \left(1 - \frac{2 a_1^2}{3 M_{\rm{Pl}}^2 g_1^2} + \frac{4\beta \phi_{\infty}}{a_1 g_1}\right)\frac{a_1 g_1^3}{4 r^4}-\bigg(1-\frac{9a_1^2}{116M_{\rm{Pl}}^2g_1^2}\nonumber\\
    &-\frac{96M_{\rm{Pl}}^2\beta g_1 \phi_{\infty}}{29a_1^3}\left[1 -\frac{3 a_1}{4 g_1 \phi_{\infty}} + \frac{a_1^2}{M_{\rm{Pl}}^2g_1^2}\right]-\frac{24M_{\rm{Pl}}^2g_1^2}{29a_1^2}\bigg)\frac{29 a_1^3 g_1^2}{120M_{\rm{Pl}}^2r^5}+\mathcal{O}(1/r^6),
\end{align}
\end{subequations}
where the following assumptions are used for the metric components and the scalar field, respectively:
\begin{align}
N^2 &= 1 + \sum_{i=1} \frac{n_i}{r^i}, \quad
g^2 = \left(1 + \sum_{i=1} \frac{g_i}{r^i}\right)^{-1}, \quad
\phi = \phi_{\infty} + \sum_{i=1} \frac{a_i}{r^i}.
\end{align}
\end{widetext}
The undetermined coefficients are identified by solving the expansion order by order in terms of $g_1$, $a_1$, and $\phi_{\infty}$, the latter corresponding to the scalar field value at infinity. 

At this point, we note that for large values of $r$, the desired asymptotic behavior is that of the Schwarzschild solution. In this context, we identify the coefficient $g_1 = -M/(4\pi M_{\rm{Pl}}^2)$, where $M$ is the ADM mass of the configuration. Therefore, if one choose a sufficiently large radius, it is possible to use the Schwarzschild metric to asymptotically estimate its mass
\begin{align}\label{Eq.DimMass}
M := \lim_{r\to\infty} 4\pi r \left(1-\frac{1}{g^2}\right) M_{\rm{Pl}}^2 \equiv \lim_{r\to\infty} m(r),
\end{align}
where in the second equality we have defined the function $m(r)$, commonly called \textit{mass function}, whose asymptotic limit as $r\to\infty$ is the ADM mass (in the asymptotically flat scenarios that we are considering). It is straightforward to verify that if we set $a_1 = 0$, then all higher-order coefficients $a_i$ are fixed to zero by the field equations, and the Schwarzschild asymptotic behavior is recover to all orders considered. In contrast, if $a_1 \neq 0$, corrections to the Schwarzschild metric appear, starting at $\mathcal{O}(1/r^3)$.

Once the boundary conditions are defined (via Eqs.~(\ref{Sys.BoundCond})), we proceed to solve the system~(\ref{Eq.MotSyst},~\ref{eq.KG.modified}) numerically using an adaptive, explicit fifth-order Runge-Kutta integration routine. The numerical error is estimated using a fourth-order method~\cite{Virtanen_2020, DORMAND198019, Shampine1986SomePR}. To facilitate the analysis, we introduce the following dimensionless variables:
\begin{align}\label{Eq.Scaling}
    \bar{\phi}=\left(\lambda_{*} M_{\rm{Pl}}\right)^{-1}\phi,\qquad \bar{r} = \lambda_{*} M_{\rm{Pl}} r,
\end{align}
where $\lambda_{*}$ is an arbitrary positive dimensionless constant. Under this scaling, the coupling parameter in Eq.~(\ref{Eq.FuncPhi}) transforms as $\bar{\beta} = \left(\lambda_{*}M_{\rm{Pl}}\right)^{2}\beta$, and the mass in Eq.~(\ref{Eq.DimMass}) becomes $\bar{M}=(\lambda_{*}/M_{\rm{Pl}})M$. Note that the dimensionless system obtained is equivalent to taking $M_{\rm{Pl}} = 1/\lambda_{*}$. To simplify the notation, we omit the bars in what follows, and use a superscript ``phys'' to denote physical (dimensionful) quantities when necessary.

Figure~\ref{fig:IllExamp} shows an illustrative configuration with mass $M = 20.1$ (equivalent to $g_1 = -1.6$), computed using $\lambda_{*} = 1$.\footnote{ Due to the scaling freedom in Eq.~(\ref{Eq.Scaling}), it is not necessary to compute solutions for different values of $\lambda_{*}$. In practice, to obtain a configuration with $\beta = \beta_{(2)}$ and $\lambda_{*} = \lambda_{(2)}$, one can compute the solution with $\lambda_{*} = 1$ and $\beta = \beta_{(2)} / \lambda_{(2)}^2$, and then rescale the results using Eq.~(\ref{Eq.Scaling}), i.e., $(r_{(2)}, \phi_{(2)}, M_{(2)}) = \lambda_{(2)} \times (r, \phi, M)$. The metric components do not require rescaling.} The boundary conditions were defined at $r_{\text{max}} = 100$ using Eqs.~(\ref{Sys.BoundCond}) with $\phi_{\infty} = 0.01$ and $a_1 = 1.0$. The left panel shows the divergence of the scalar field at the origin, while the Ricci and Kretschmann scalars approach the analytical values given by Eqs.~(\ref{Eq.ScCurvOrig}). From the right panel, we conclude that although $a_1 \neq 0$, the corrective contributions for $r > r_{\text{max}}$ are negligible compared to the Schwarzschild solution. On the other hand, note that although we solve the system~(\ref{Eq.MotSyst},~\ref{eq.KG.modified}) from $r_{\text{max}} = 100$ towards the origin, the numerical solution for the $g_{rr}$ metric component naturally satisfies $g_{rr}(r = 0) = 1$, ensuring finite curvature. In other words, no fine-tuning of parameters is required. Notice that we also integrate outward from $r_{\text{max}} = 100$ up to $r = 3 \times r_{\text{max}}$.

Figure~\ref{fig:IllExamp2} presents a comparison between the numerical profiles for the scalar field (left panel), the $g_{tt}$  (center panel) and $g_{rr}$ metric components (right panels), and their corresponding approximate solutions near the origin Eqs.~(\ref{Sys.OrigCond}) and at infinity Eqs.~(\ref{Sys.BoundCond}). As observed, the agreement with Eqs.~(\ref{Sys.OrigCond}) is limited to a small region near the origin, whereas Eqs.~(\ref{Sys.BoundCond}) provide a more accurate description for $r > r_{\text{max}}$. In fact, although not shown in the figure, they can also describe regions with $r < r_{\text{max}}$.

%%%%%%%%%%%%%%%%%%%%%%%%
\subsection{Hybrid Einstein-scalar-Gauss-Bonnet stars}\label{EGBHS}
%%%%%%%%%%%%%%%%%%%%%%%%

Here we consider the case where fermionic matter is present, i.e., $\epsilon \neq 0$, $p \neq 0$. For this case, as discussed in subsection~\ref{SubSectFinitCurvOrig}, regularity at the origin is ensured. In addition to the system of equations~(\ref{Eq.MotSyst},~\ref{eq.KG.modified}), we must include the equation of hydrostatic equilibrium of the fluid. This equation can be derived from the condition\footnote{For the metric~(\ref{metric}), using the scalar field equation~(\ref{eq.KG.modified}), one can show that $\nabla^{\mu}G^{(\rm{GB})}_{\mu\nu} = \nabla^{\mu}T_{\mu\nu}^{(\phi)}$, which leads to $\nabla^{\mu}T_{\mu\nu} = 0$.}
\begin{equation}\label{Eq.Conserv}
\nabla^{\mu}T_{\mu\nu} = 0 \quad \Rightarrow \quad p' = - (\epsilon + p)\frac{N'}{N},
\end{equation}
where Eq.~(\ref{perfl}) and the dimensionless variables~(\ref{Eq.Scaling}) are used. The pressure $p$ and total energy density $\epsilon$ are scaled as:
\begin{align}
   \bar{p}=\left(\lambda_{*} M_{\rm{Pl}}\right)^{-4} p, \quad \bar{\epsilon} = \left(\lambda_{*} M_{\rm{Pl}}\right)^{-4} \epsilon.
\end{align}
Note that, following our notation, we omit the bar in Eq.~(\ref{Eq.Conserv}).

To close the system of equations we need to write down an equation of state for the fluid component, for which we consider the (simple) polytropic case~\cite{Roque:2021lvr}
\begin{align}
    p=k\rho^{\Gamma},
\end{align}
where $\Gamma$ is the adiabatic index, fixed in this work to $\Gamma=2$; $k$ is the polytropic constant, set to $k^{\text{phys}}=100 \left(\lambda_{*} M_{\rm{Pl}}\right)^{-4}$; and $\rho$ is the mass density. These quantities are related to the energy density by:
\begin{align}\label{Eq.EOS}
    \epsilon=\left(\frac{p}{k}\right)^{1/\Gamma}+\frac{p}{\Gamma-1}.
\end{align}
It is worth noting that we did not use realistic equations of state (e.g., SLy2/4~\cite{CHABANAT1998231, 2001A&A...380..151D}, FPS~\cite{FRIEDMAN1981502, PhysRevLett.70.379}, or others~\cite{composeWeb}) in our analysis, in order to avoid fixing the value of $\lambda_{*}$ and to keep the results as general as possible. The objective of this work is to present these novel solutions and to identify their main characteristics. In future work, we will use observational astrophysical data to constrain the value of $\lambda_{*}$. (Recall that $\lambda_{*}$ is related to the coupling constant $\beta$.)

%%%%%%%%%%%%%%%%%%%%%%%%%%%%
\subsubsection{Regular scalar field at the origin}\label{Sec.RegScalFieldatOrig}
%%%%%%%%%%%%%%%%%%%%%%%%%%%%
\begin{figure*}%\label{fig:WCValidity}
    \centering
    \includegraphics[width=1.\linewidth]{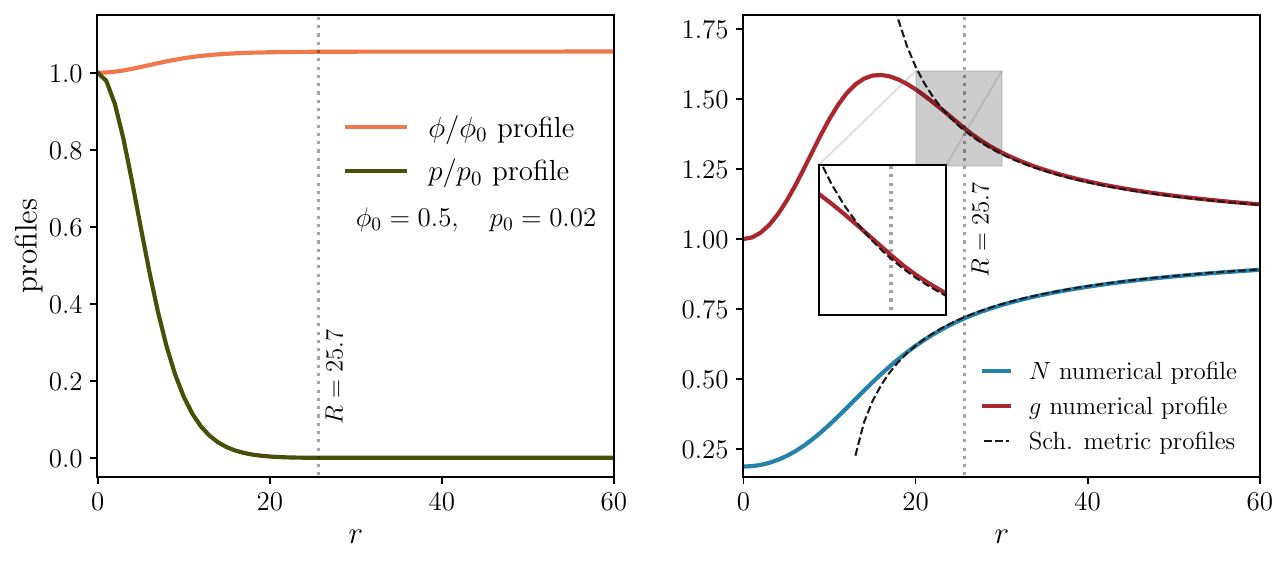}
    \caption{\textbf{Illustrative Einstein–Gauss–Bonnet hybrid stars with a regular scalar field at the origin.} Left panel shows the matter pressure and scalar field profiles, scaled by their respective central values: $p_0=0.02$ and $\phi_0=0.5$ with $\beta=1.0$. Right panel displays the metric function: lapse $N(r)$ and radial component $g(r)$, both normalized to one at spatial infinity. The dashed lines represent the Schwarzschild metric components for an object of the same total mass, $M=155$ (see Eq.~(\ref{Eq.DimMass})). The vertical lines indicate the radii where $p= 0$ (border of the baryonic component of the star).}\label{fig:MIllExamp}
\end{figure*}

To get the desired hybrid stars profiles, we need to solve numerically the system~(\ref{Eq.MotSyst},~\ref{eq.KG.modified}) together with the hydrostatic equation~(\ref{Eq.Conserv}), and the equation of state~(\ref{Eq.EOS}). For this, it is necessary to define a set of boundary conditions in the center that correspond to regular spacetime configurations (no divergences of curvature scalars), with a non-divergent scalar field. As discussed in subsection~\ref{SubSectFinitCurvOrig}, for this case we have:
\begin{align}\label{Eq.BoundCond}
& g(r=0)=1, \qquad g'(r=0)=0, \nonumber\\
& N(r=0)=N_0, \quad N'(r=0)=0, \\
& \phi(r=0)=\phi_{0}, \;\;\quad \phi'(r=0)=0, \quad p(r=0)=p_0, \nonumber
\end{align}
where $\phi_{0}$ is the field amplitude at the origin, $p_0 $ is the central fermionic pressure of the star, and $N_0$ the lapse function evaluated at the center of the configuration. These are free and positive constants that one can choose arbitrarily.\footnote{One can use the freedom to rescale the metric component, $N \to \lambda N$, with $\lambda$ an arbitrary constant, to fix $N_{0}=1$. Later, from the asymptotic value $N_{\infty}$, one can determine the value of $\lambda$ such that $N_\infty = 1$, and thereby identify the appropriate value of $N_{0}$.} At this point, it is important to note that, in practice, we do not integrate from $r = 0$, but rather from $r = r_{\text{min}} \ll 1$, where the boundary conditions are defined through a Taylor series expansion of the functions $p$, $g$, $N$, and $\phi$. To determine the expansion coefficients, it is necessary to consider the real roots of the system~(\ref{Eq.OrigSecDeriva}).

Figure~\ref{fig:MIllExamp} shows an illustrative example of these types of Einstein–Gauss–Bonnet-hybrid stars with $\beta=1$, in this case characterized by $\phi_0 = 0.5, p_0 = 0.02$. As noted in the left panel, the pressure drops to zero at the baryonic radius $R \approx 25.7$, while the scalar field settles to a non-zero constant. One could consider the solution to be non-physical, since the scalar field does not decay exponentially as $r \to \infty$.\footnote{ In Ref.~\cite{Ventagli:2021ubn}, the authors study similar objects, but with the inclusion in the action~(\ref{eq:lag}) of an additional term, $\alpha R/2$, coupled to the scalar field. They report the existence of solutions where $\lim_{r \to \infty} \phi = 0$. However, it is not clear to us whether their results include the case $\alpha = 0$ (which corresponds to our setup). In fact, from the system~(\ref{Eq.OrigSecDeriva}), one can show that when $\phi_0'' < 0$ (which implies $\phi'(r_{\text{min}}) < 0$ at first order), then $g_0'' < 0$, which is not observed in the solutions. See Appendix~\ref{Ap.Demostrat} for an intuitive demonstration.} However, as seen in the right panel, it is only near the baryonic radius that one can distinguish between this object and the Schwarzschild solution. In other words, at $r \to \infty$, we cannot discern between general relativity and Einstein–Gauss–Bonnet solutions. We will return to this discussion in Section~\ref{Sec.Phenomenology}.

%%%%%%%%%%%%%%%%%%%%%%%%%%%%
\subsubsection{Non-regular scalar field at the origin}
%%%%%%%%%%%%%%%%%%%%%%%%%%%%
\begin{figure*}
    \centering
    \includegraphics[width=1.\linewidth]{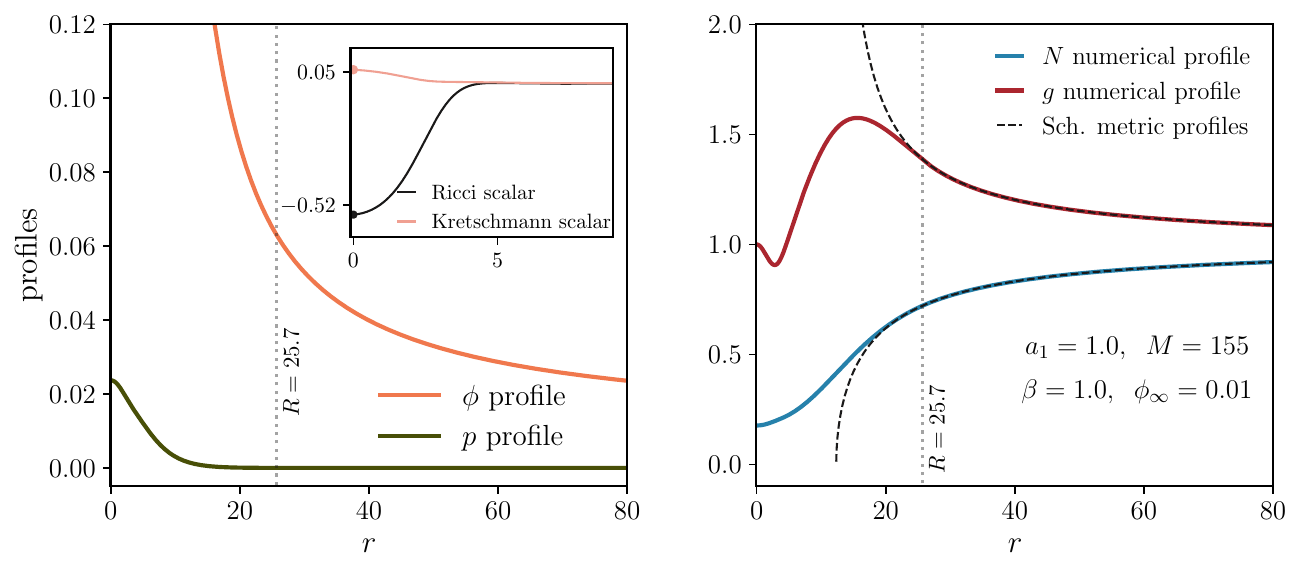}
    \caption{\textbf{Illustrative Einstein–Gauss–Bonnet hybrid stars with a non-regular scalar field at the origin.} The configuration is equivalent to that shown in Fig.~\ref{fig:MIllExamp} in the sense that it has the same total mass $M = 155$ and baryonic radius $R = 25.7$. However, its metric, pressure, and scalar field profiles exhibit significant differences. The inset displays the Ricci and Kretschmann scalar profiles. In the right panel, we show the values used to determine the boundary conditions from the asymptotic expansion~(\ref{Sys.BoundCond}).}\label{fig:MIllExamp2}
\end{figure*}

A natural question arises: do stellar configurations analogous to those presented in Section~\ref{EGBSS} exist, but with a fermionic matter component? To explore this, we developed a methodology similar to that used in the vacuum case:
\begin{enumerate}
    \item[i.] First, at a sufficiently large radius $r_{\text{max}}$, where the pressure vanishes ($p = 0$), we impose boundary conditions derived from the asymptotic expansion~(\ref{Sys.BoundCond}) and numerically solve the system~(\ref{Eq.MotSyst},~\ref{eq.KG.modified}) in the vacuum region, integrating inward to a chosen baryonic radius $R \ll r_{\text{max}}$.
    \item[ii.] Second, at the baryonic radius $R$, we introduce the matter pressure, adopt the metric and scalar field boundary conditions obtained in the previous step, and solve the system~(\ref{Eq.MotSyst},~\ref{eq.KG.modified}) along with the hydrostatic equilibrium equation~(\ref{Eq.Conserv}) and the equation of state~(\ref{Eq.EOS}) towards the origin.
    \item[iii.] Finally, using the complete metric profiles, we compute the Ricci and Kretschmann scalars.
\end{enumerate}

Figure~\ref{fig:MIllExamp2} shows an illustrative configuration ``equivalent'' to Fig.~\ref{fig:MIllExamp} (in the sense that it has the same total mass $M = 155$, baryonic radius $R = 25.7$ and $\beta=1.0$), but now, the scalar field diverges at the origin. Left panel presents both the scalar field and pressure profiles. Note that both decrease monotonically; however, the pressure remains finite at $r = 0, p_0\approx 0.024$, while the scalar field does not. The inset displays the Ricci and Kretschmann scalar profiles. As observed, their values remain finite at all radii, and the values at the origin match those predicted by Eqs.~(\ref{Eq.ScCurvOrig}). It is also worth noting that the Ricci scalar approaches zero asymptotically from negative values, indicating that the spacetime is negatively curved—that is, the geometry is ``open''. Such curvature may arise from particular energy-momentum configurations, potentially involving negative pressure (as in dark energy models) or vacuum solutions with a negative cosmological constant (e.g., anti-de Sitter space). The right panel shows a comparison between the metric profiles and an equivalent Schwarzschild solution. An atypical behavior of the metric components inside the star is evident. In the next section, we discuss possible physical implications of this feature.

%%%%%%%%%%%%%%%%%%%%%%%%
\section{Main phenomenological characteristics}\label{Sec.Phenomenology}
%%%%%%%%%%%%%%%%%%%%%%%%

In the previous section, we demonstrated the numerical existence of static, spherically symmetric solutions within the framework of Einstein-scalar-Gauss-Bonnet gravity. We classified these compact objects into two categories---scalar stars and hybrid stars---and showed that certain configurations admit a divergent scalar field at the origin while still maintaining finite scalar curvature. In what follows, we present and analyze the main phenomenological characteristics of these solutions.

%%%%%%%%%%%%%%%%%%%%%%%%
\subsection{Mass–Radius relationship}
%%%%%%%%%%%%%%%%%%%%%%%%
\begin{figure*}
\centering
\includegraphics[width=1.\linewidth]{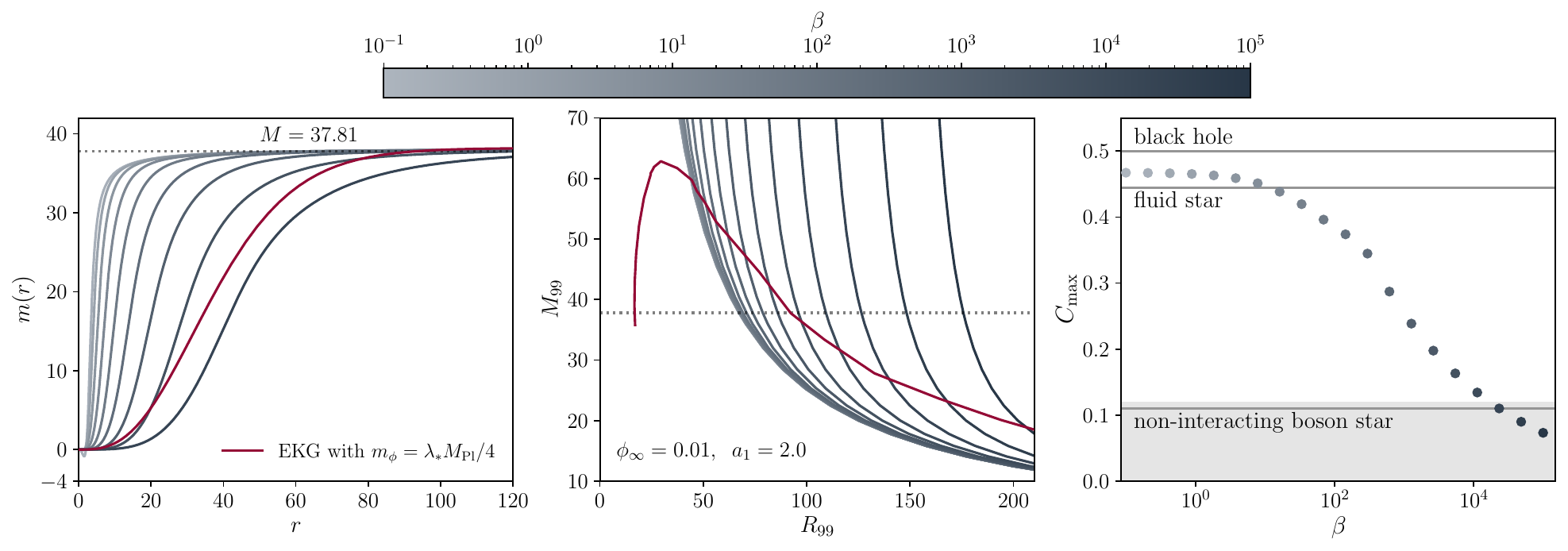}
\caption{\textbf{Mass–Radius relationship of Einstein–Gauss–Bonnet scalar stars.} The left panel shows the mass-radius profiles (computed using Eq.~(\ref{Eq.DimMass})) corresponding to selected configurations with dimensionless mass $M = 37.81$ (dotted line) displayed in the central panel. Note that for $\beta \sim 10^{-1}$, the mass becomes negative near the origin---an atypical behavior discussed in the main text. The center panel shows the mass $M_{99}$ as a function of the effective radius $R_{99}$ for configurations with $\phi_{\infty}=0.01$ and $a_1=2.0$. Each curve corresponds to a different value of $\beta$ in the range $[10^{-1}, 10^{5}]$. For comparison we also show the Einstein-Klein-Gordon family of solutions (red curve). The right panel presents the maximum compactness $C_{\textrm{max}}$ (computed using Eq.~(\ref{Eq.Compactness})) as a function of $\beta$.  We also display the following maximum compactness limits: Schwarzschild black hole ($C=1/2$),  Buchdahl limit for fluid stars ($C_{\textrm{max}}=4/9$)~\cite{Buchdahl:1959zz}, and boson star without self-interactions ($C_{\textrm{max}}=0.1$). Configurations lying above the shaded region correspond to super-emitters, in the terminology of Ref.~\cite{Palenzuela:2017kcg}. For further details, see the discussion in Section~\ref{SectMR_sub_Ss}.}\label{fig:scalarMass}
\end{figure*}

The main characteristics of compact objects are defined in terms of their mass–radius relationship.\footnote{ We define the effective radius of the object, $R_{99}$, as the radius within which $99\%$ of the total mass $M_T$ is contained; that is, $M_{99} = 0.99 M_T$.} This relation provides essential information for classifying the gravitational regime of the object---whether Newtonian, weak-field, or strong-field---as well as for validating stellar evolution models, calculating orbital parameters, and predicting the future dynamics of binary systems. In what follows, we present our results on this relationship in the context of Einstein–Gauss–Bonnet gravity.

%%%%%%%%%%%%%%%%%%%%%%%%
\subsubsection{Scalar stars}\label{SectMR_sub_Ss}
%%%%%%%%%%%%%%%%%%%%%%%%

Figure~\ref{fig:scalarMass} summarizes the main results for the mass relation of pure Einstein-scalar-Gauss-Bonnet stars. As discussed in Section~\ref{EGBSS}, these objects exist only within this gravitational model and are characterized by a non-regular scalar field at the origin, while still remaining regular compact objects. The left panel shows illustrative profiles of $m(r)$ (computed using Eq.~(\ref{Eq.DimMass})) corresponding to a set of values of $\beta$ in the range $[10^{-1}, 10^{5}]$. We see that for smaller values of the coupling parameter, $\beta \sim \mathcal{O}(10^{-1})$, these profiles exhibit atypical behavior compared to the usual profiles found in general relativity (denoted by a red color), taking negative values near the center. Although not shown in the left panel, when $\beta \sim \mathcal{O}(10^{4})$ and the total mass is on the order of $200$, the profile reaches a maximum before decreasing toward the asymptotic mass value. These atypical behaviors highlight that, although 
$m(r)$ is sometimes referred to as the mass function, its interpretation as the mass of the stellar configuration is justified only in the asymptotic limit.

From the left and center panels of Fig.~\ref{fig:scalarMass}, it is evident that increasing the coupling parameter $\beta$ leads to more extended mass profiles and consequently, to larger objects (i.e., higher values of $R_{99}$). In particular, for sufficiently large $\beta$--as seen in the configurations located to the right of the red curve in the center panel--these scalar stars become even more extended than their non-interacting boson star counterparts with scalar mass $m_{\phi}=\lambda_{*}M_{\rm{Pl}}/4$. Moreover, the center panel shows that, unlike compact objects in general relativity, these scalar stars do not exhibit a maximum‐mass configuration (at least for the mass range studied, on the order of $200{\text{--}}300$). The absence of a turning point in the mass curve renders the usual stability criterion inapplicable---at least as a first approximation. A dedicated stability analysis of these objects will be presented in future work. Finally, we note that the choice of the asymptotic scalar field value, $\phi_{\infty}$, has little impact on these results. In contrast, variations in the parameter $a_{1}$ do affect the detailed shape of the mass profiles, though not the qualitative conclusions discussed above.\footnote{ The roles of the quantities $\phi_{\infty}$ and $a_{1}$ can be partially inferred from their units: $\phi_{\infty}$ is proportional to the Planck mass, $M_{\rm{Pl}}$, while $a_{1}$ is a dimensionless parameter.}

The right panel of Fig.~\ref{fig:scalarMass} shows the maximum compactness, $C_{\text{max}}$, allowed for these scalar stars as a function of the coupling parameter $\beta$, for the fixed choices $\phi_{\infty} = 0.01$ and $a_1 = 2.0$.\footnote{By analogy with stellar objects, we define the compactness as the ratio between 99\% of the total mass of the configuration and the radius that encloses that mass, namely:
\begin{align}
C = \frac{M_{99}}{8\pi M_{\rm{Pl}}^2 R_{99}}.\label{Eq.Compactness}
\end{align}
}
Note that for coupling values $\beta \lesssim 10^{4}$, we identify configurations with compactness exceeding the maximum limit for non-self-interacting boson stars, $C\approx0.1$~\cite{Liebling:2012fv}. More generally, we find scalar star configurations whose compactness surpasses that obtained by including an attractive $\lambda \phi^4$ self-interacting term, or even the highest compactness values reported in the literature--around $C\approx 0.33$-- for various scalar field potentials~\cite{PhysRevD.35.3658, Cardoso:2016oxy, Palenzuela:2017kcg}. These compactness value also exceed those typically associated to neutron stars $C\approx 0.1-0.2$~\cite{Xtreme, Haensel:2007yy}, and in some cases is possibles exceeding the theoretical upper bound for fluid stars, the Buchdahl limit $C=4/9$~\cite{Buchdahl:1959zz}. However, all studied configurations remain below the compactness of a Schwarzschild black hole, $C=1/2$.

Finally, to close this subsection, we briefly discuss the potential gravitational radiation emittance resulting from the collision of two such objects. For this purpose, we adopt a simplified model~\cite{Braganca:2018rnm, Palenzuela:2017kcg, Barranco:2021auj}\footnote{ This estimate was originally derived in a general relativistic context~\cite{PhysRevD.35.3658, Cardoso:2016oxy, Palenzuela:2017kcg, Hanna:2016uhs}, but the assumptions are sufficiently general that, with appropriate caution, Eq.~(\ref{eq:xirad}) may be applied within the framework of Einstein–Gauss–Bonnet gravity.}
\begin{equation}
\xi_\textrm{{rad}} \approx 0.48 C M_S , \label{eq:xirad}
\end{equation}
where $M_S=M_{T_1}+M_{T_2}=2M_T$ is the total initial mass of the binary system ($T_1, T_2$), and $C$ is the compactness of the objects, which is considered to be the same. Although we do not carry out a full dynamical analysis, some insights can be drawn based on the compactness of the configurations. Using~\eqref{eq:xirad} as an upper bound--and noting that binary black hole mergers radiate approximately $5\%$ of their total mass in gravitational waves~\cite{Pollney:2009yz, Abbott:2016blz}--it is conceivable that these scalar stars could emit even more, provided their compactness exceeds $C \gtrsim 0.12$. Configurations above the shaded region in the right panel of Fig.~\ref{fig:scalarMass} reach a compactness higher than this threshold, which means that they potentially qualify as super-emitters in the terminology of~\cite{Palenzuela:2017kcg}.

%%%%%%%%%%%%%%%%%%%%%%%%
\subsubsection{Hybrid stars}
%%%%%%%%%%%%%%%%%%%%%%%%

\begin{figure*}
    \centering
    \includegraphics[width=1.\linewidth]{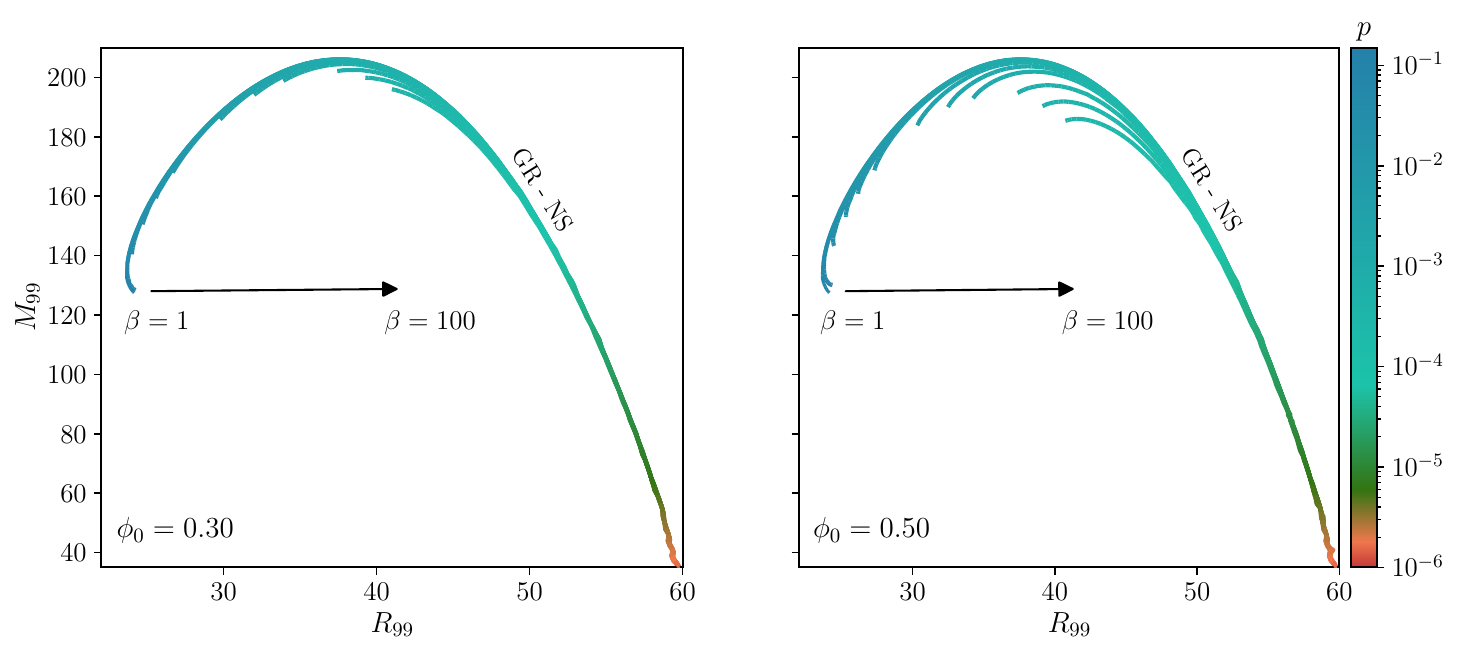}
    \caption{\textbf{Mass–Radius relationship of Einstein-Gauss-Bonnet hybrid stars (with a regular scalar field).} Each curve corresponds to a set of configurations with $\beta \in [1, 3, 5, 7, 10, 20, 30, 40, 60, 80, 100]$ (increasing as indicated by the arrow), computed for varying the central pressure $p_0$ in the range $[10^{-6}, 10^{-1}]$ (vertical gradient color bar), for two values of the scalar central amplitude: $\phi_0=0.30$ (left panel) and $\phi_0=0.50$ (right panel). Note that the neutrons stars in general relativity (GR-NS) represent the upper limits of the configurations. The fermionic matter is described by Eq.~(\ref{Eq.EOS}) with $\Gamma=2$ and $k=100$. See the text for more details.}\label{fig:barionic_regularMass}
\end{figure*}

\begin{figure*}
    \centering
    \includegraphics[width=1.\linewidth]{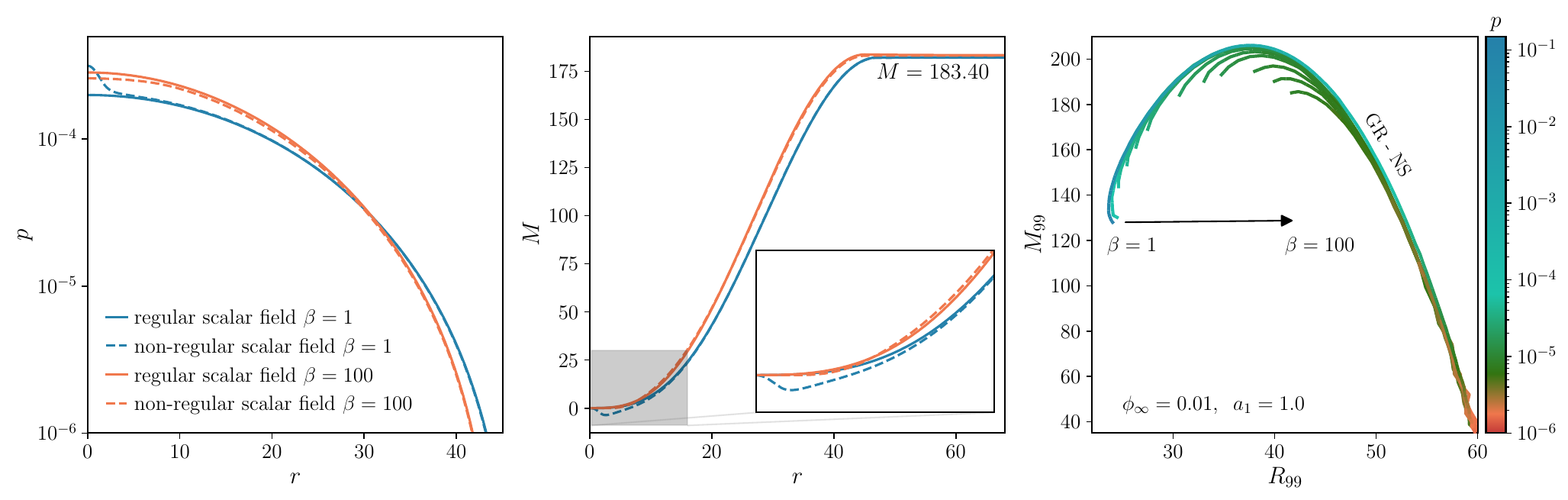}
    \caption{\textbf{Pressure–Mass–Radius relationship of Einstein–Gauss–Bonnet hybrid stars (with a non-regular scalar field)}. Using the mass and baryonic radius corresponding to the right panel of Fig.~\ref{fig:barionic_non_regularMass}, and setting $\phi_{\infty} = 0.01$ and $a_1 = 1.0$, we identify a family of configurations with $\beta \in [1, 3, 5, 7, 10, 20, 30, 40, 60, 80, 100]$. The Mass–Radius relationship is shown in the right panel, while the center and left panels display the mass and pressure profiles corresponding to $\beta, M = (1, 183.4)$ and $(100, 183.4)$, respectively. These results are discussed in the main text.}\label{fig:barionic_non_regularMass}
\end{figure*}

In section~\ref{EGBHS}, we prove the theoretical existence of hybrid Einstein-scalar-Gauss-Bonnet stars. In particular, we identify two branches: one that is connected in some limit with its equivalent in general relativity, and another one that is not. The first branch is associated with a regular scalar field at the origin, while the second branch has a non-regular scalar field.

Figure~\ref{fig:barionic_regularMass} shows the mass-radius relation for a set of hybrid stars with $\beta \in [1, 3, 5, 7, 10, 20, 30, 40, 60, 80, 100]$, all having a regular scalar field at the origin. The central amplitudes are: $\sigma_0 = 0.30$ (left panel) and $\sigma_0 = 0.50$ (right panel). As can be seen, the equivalent neutron stars (denoted as GR-NS) represent the upper mass bound at a fixed radius. When the value of $\beta$ increases, the masses of the configurations become smaller than those of the corresponding GR-NS. This can be understood as the Gauss-Bonnet terms contributing ``negatively'' to the gravitational forces, resulting in a lower mass (compared to the case where these terms are absent) being required to achieve hydrostatic equilibrium. It is important to note that the central pressures are of the same order. This result is consistent with that reported in Ref.~\cite{Roque:2021lvr} for the Horndesky hybrid star family of solutions.

As expected, we can conclude from Fig.~\ref{fig:barionic_regularMass} that when the central amplitude $\phi_0 \to 0$ and/or $\beta \to 0$, the configurations approach their counterparts in the general relativity limit.\footnote{In the limiting case where fermionic matter dominates ($\phi_0 \sim 0$), we find $R_{99} \sim R$, where $R$ is the radius at which $p(R) = 0$.} Notice that, qualitatively, a small value of the central amplitude suppresses the modifications introduced by the coupling parameter. This behavior can be understood in terms of Eq.~(\ref{Eq.Scaling}): $\phi$ is modulated by $\lambda M_{\rm{Pl}}$, while $\beta$ is modulated by $(\lambda M_{\rm{Pl}})^{-2}$. Finally, the compactness of these hybrid stars is lower than that of the corresponding general relativity neutron stars.

The right panel of Fig.~\ref{fig:barionic_non_regularMass} reproduces the mass–radius relation shown in the same panel of Fig.~\ref{fig:barionic_regularMass}, but this time considering configurations with a non-regular scalar field (i.e., $\lim_{r \to 0} \phi \to \infty$). For this example, we set $\phi_{\infty} = 0.01$ and $a_1 = 1.0$, and extracted the mass and baryonic radius from the previously discussed Figure. The results reveal both similarities and differences between these types of stars, as we explain below.\newline

\noindent \textit{Similarities:} Configurations with a non-regular scalar field are capable of mimicking the previous (regular) case, with general relativity still setting the upper mass limit. The mass profiles at large radii exhibit the same behavior (see the central panel of Fig.~\ref{fig:barionic_non_regularMass}), despite the scalar fields having different asymptotic behaviors.
\newline

\noindent \textit{Differences:} In this case, the general relativity limit $\beta \to 0$ does not exist (the limit $\phi_0 \to 0$ is not viable because $\phi_0 \to \infty$), as demonstrated in Section~\ref{SubSectFinitCurvOrig}. This can be inferred from the left and central panels of Fig.~\ref{fig:barionic_non_regularMass}. As can be observed, when $\beta = 1$, some atypical behaviors appear near the origin ($r = 0$) compared to the regular case (solid lines). The pressure drops sharply, and the mass profile becomes negative. These behaviors are directly related to the fact that, as $\beta \to 0$, the curvature scalar diverges at the origin (see Eq.~(\ref{Eq.ScCurvOrig})). Another difference is that for values of $\beta$ sufficiently far from zero, the central pressure required to mimic the same configuration as in the regular case is lower (see the left and right panels of Fig.~\ref{fig:barionic_non_regularMass}). This is particularly interesting, as it could help in understanding the existing tension between equations of state and gravity models when considering tidal deformability measurements from binary neutron star mergers~\cite{Hinderer:2009ca, De:2018uhw, LIGOScientific:2018cki, Chatziioannou:2020pqz, Biswas:2023ceq}.

%%%%%%%%%%%%%%%%%%%%%%%%
\subsection{Astrophysical classification via \texorpdfstring{$\lambda_{*}$}{lambda*} scaling}
%%%%%%%%%%%%%%%%%%%%%%%%
\begin{figure*}%\label{fig:WCValidity}
    \centering
    \includegraphics[width=1.\linewidth]{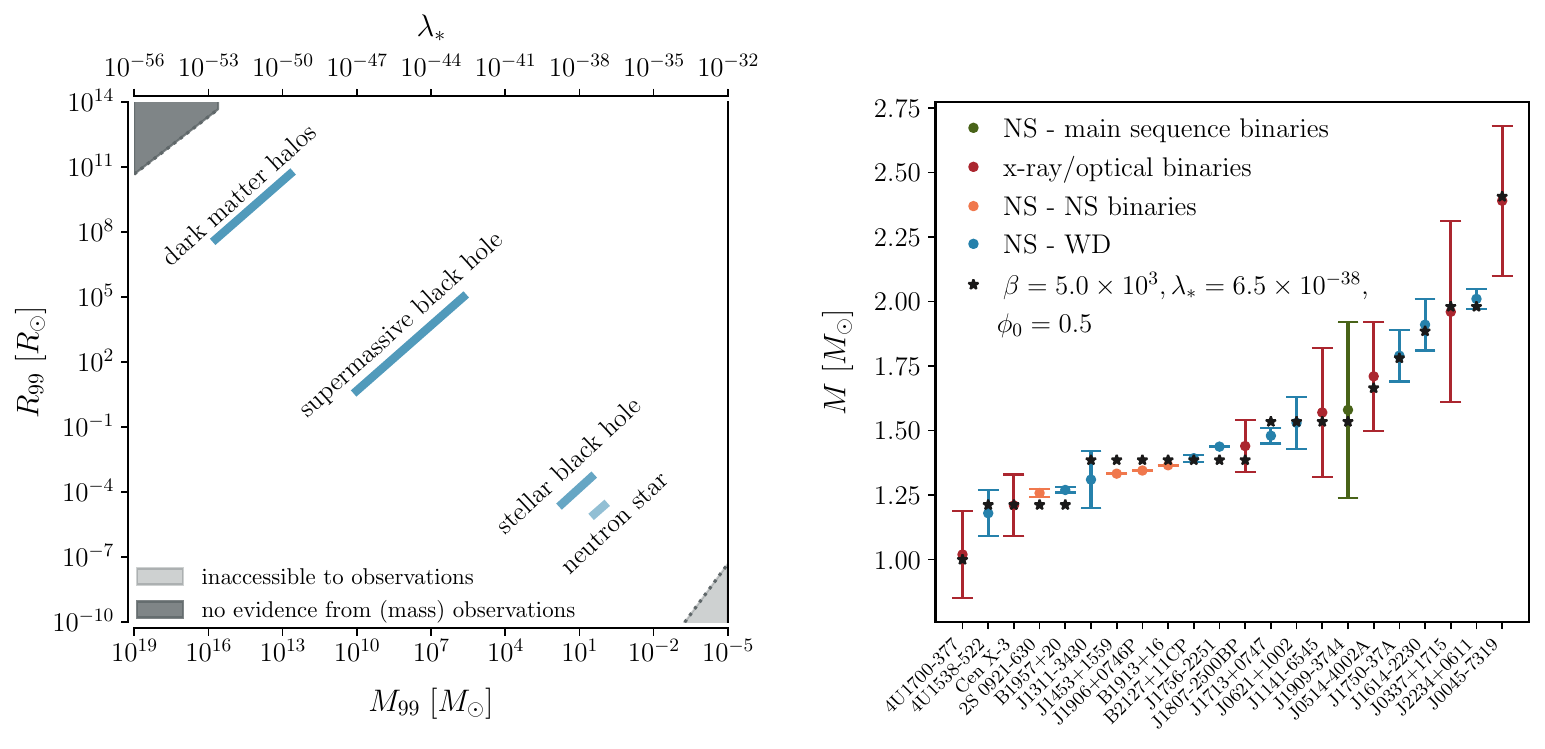}
    \caption{\textbf{Astrophysical classification.}  Using the relations in Eq.~(\ref{Eq.PhysU}), the left panel shows regions (blue lines) associated with the $\lambda_{*}$ values (upper axes) that are capable of reproducing the mass and radius corresponding to different astrophysical compact objects. We used as a reference the typical values of an Einstein–Gauss–Bonnet scalar star with $M_{99} = 37.81$ and $R_{99} = 90$. The right panel shows that a family of hybrid stars with $\phi_0 = 0.5$ and $\beta = 5 \times 10^3$ (denoted by star markers) can reproduce the observed masses of neutron stars (circular markers) when $\lambda_{*} \sim 10^{-38}$.}\label{fig:physReg}
\end{figure*}

Now we proceed to classify, in terms of the constant $\lambda_{*}$ introduced in Eq.~(\ref{Eq.Scaling}), the astrophysical objects that could be represented by Einstein-scalar-Gauss-Bonnet stars. 

The physical units are recovered by means of:
\begin{subequations}
\begin{align}\label{Eq.PhysU}
    M^{phys} &= \frac{M_{\rm{Pl}}}{\lambda_{*}} M = 1.09 M \times \frac{10^{-38}}{\lambda_{*}}M_{\odot}, \\
    R^{phys} &= \frac{\hbar}{\lambda_{*} c M_{\rm{Pl}}} R = 2.32 R \times \frac{10^{-44}}{\lambda_{*}}R_{\odot},
\end{align}
\end{subequations}
where $M_{\odot}\approx 1.99\times 10^{30}$ kg~\cite{2016AJ....152...41P} and $R_{\odot}\approx 6.96\times 10^{5}$ km~\cite{2012ApJ...750..135E, 2016AJ....152...41P} are the mass and radius of Sun respectively.

Considering a typical pure Einstein-scalar-Gauss-Bonnet star with $M_{99} = 37.81$ and $R_{99} = 90$ (see Fig.~\ref{fig:scalarMass}), we find that when the constant $\lambda_{*} \sim 10^{-37}$, these stars exhibit phenomenological characteristics—i.e., mass and radius—comparable to those of typical neutron stars. Similarly, stellar and supermassive black holes can be reproduced for $\lambda_{*} \sim 10^{-38}$ and $\lambda_{*} \sim 10^{-45}$, respectively. For $\lambda_{*} \sim 10^{-51}$, the properties of these stars resemble those of dark matter halos. In fact, for comparison with typical boson stars, if we take the scalar field mass to be $m_{\phi} = \lambda_{*} M_{\textrm{Pl}}$, then for $\lambda_{*} \sim 10^{-45}$, the corresponding value is of the order of the ultralight scalar dark matter mass, $m \sim 10^{-23}$ eV~\cite{Gonzalez-Morales:2016yaf, Zimmermann:2024xvd, Teodori:2025rul}. The left panel of Fig.~\ref{fig:physReg} summarizes the above discussion, outlining the different regions in parameter space with a blue line.

In the right panel of Fig.~\ref{fig:physReg}, we demonstrate that even for a polytropic equation of state Eq.~(\ref{Eq.EOS}), hybrid Einstein-scalar-Gauss-Bonnet stars are capable of reproducing the observed masses of neutron stars. Using an observed neutron star mass catalog Refs.~\cite{catalgNeutronStar, 2012ARNPS..62..485L}, we randomly selected a set of objects whose masses were measured using different techniques (with distinct markers used to indicate each technique), along with a family of hybrid stars corresponding to $\phi_0 = 0.5$ and $\beta = 5.0 \times 10^{3}$ (denoted with a star marker). By setting $\lambda_{*} = 6.5 \times 10^{-38}$, we were able to reproduce each configuration with acceptable confidence. It is important to note that the radii of these hybrid stars are on the order of the expected neutron star radius---approximately $10-14$ km---and that, as a consequence of the \textit{similarities} with the hybrid Einstein-scalar-Gauss-Bonnet stars with non-regular scalar fields, these results are equally valid for them.

%%%%%%%%%%%%%%%%%%%%%%%%
\subsection{Trajectories of massive particles coupled to the scalar field}
%%%%%%%%%%%%%%%%%%%%%%%%
\begin{figure*}
    \centering
    \includegraphics[width=1.\linewidth]{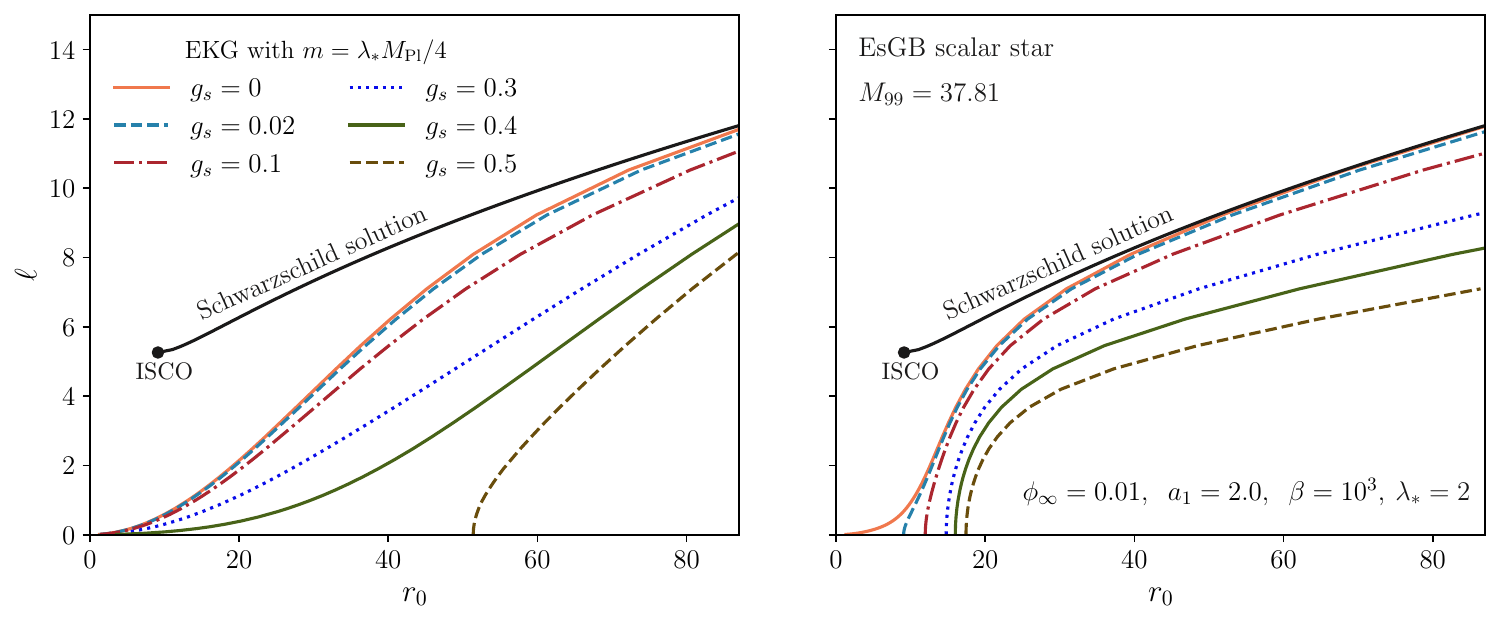}
    \caption{\textbf{Stable circular orbit family as a function of the coupling parameter $g_s$.} The figure shows the relationship between the angular momentum $\ell$ and the corresponding stable circular orbit radius $r_0$ for a range of values of the scalar–matter coupling parameter $g_s$. The configurations used in both panels have a fixed mass $M_{99} = 37.81$ (see Fig.~\ref{fig:scalarMass}). The left panel corresponds to an Einstein–Klein–Gordon (EKG) star, while the right panel shows results for an Einstein–Gauss–Bonnet (EsGB) scalar star. To enable a direct comparison between the two configurations, we fix on the EKG theory the scalar mass $m = \lambda_{*} M_{\textrm{Pl}}/4$ and $\lambda_{*} = 2$. The specific parameter values used are indicated within the plots. All displayed values of $r_0$ satisfy the conditions in Eqs.~(\ref{Eq.PotCond},~\ref{Eq.Condit}), representing the locations of the minima of the effective potential~(\ref{eq:weff}) corresponding to different values of angular momentum.}\label{fig:lvsrmin}
\end{figure*}

In the previous sections, we studied static, spherically symmetric and asymptotically flat spacetimes in the presence of the additional scalar degree of freedom that arises from the non-minimal coupling between the scalar field and the Gauss–Bonnet term. We identified hybrid star solutions in which this scalar degree of freedom is minimally coupled to the matter sector—specifically, to a fermionic field. In what follows, we explore the impact of introducing a non-minimal coupling between the scalar field and the matter sector, focusing in particular on its effect on the innermost stable circular orbit.

The study of particle motion provides insights on the properties of spacetime. In metric theories of gravity---and in the absence of non-gravitational interactions---the trajectories of freely falling test bodies are geodesics of the spacetime metric. However, this is no longer the case when additional interactions are present. For example, charged particles in an electromagnetic field experience a Lorentz force and deviate from geodesic motion.

In the presence of a scalar field, there is no unique prescription for coupling massive particles to the scalar, and various formulations have been proposed in the literature (see, e.g., Refs.~\cite {Misner:1972jf, Breuer:1973ktj, Dahia:2007ep, Braganca:2018rnm, Kuntz:2019zef, Stashko:2022dtx, Turimov:2024orr}). In this work, we adopt the approach introduced in Refs.~\cite{Misner:1972jf, Breuer:1973ktj}. The total action is given by
\begin{widetext}
\begin{align} \label{eq:lagcp}
S =\int d^{4}x \sqrt{-g}\bigg(\frac{M_{\textrm{Pl}}^2}{2}R-X+\gamma f(\phi)\mathcal{L}_{GB}\bigg) -  m 
\int d\lambda \left(1+g_s \frac{\phi}{M_{\textrm{Pl}}}\right) \sqrt{-u_\mu u^\mu}\,, 
\end{align}
\end{widetext}
where $m$ is the bare (non-geometrized) mass of the point particle, $u^{\mu}=dz^{\mu}/d\lambda$ is the unit timelike four-velocity of the particle, $\lambda$ parametrizes the particle's worldline $z^{\mu}(\lambda)$, and $g_s$  is a dimensionless coupling constant associated with the scalar ``charge'' of the particle. Note that the presence of the scalar field modulates the particle's effective mass according to: $m_{\text{eff}}=m(1+g_s \phi/M_{\textrm{Pl}})$.

With this Lagrangian~(\ref{eq:lagcp}), the metric field equations~\eqref{met_eq} are modified by the addition of a term $T^{(m\phi)}_{\mu\nu}/M_{\textrm{Pl}}^2$ on the right hand side, where 
\begin{align}
T^{(m\phi)}_{\mu\nu}= -\frac{m}{\sqrt{-g}}\int d\lambda \left(1+g_s \frac{\phi}{M_{\textrm{Pl}}}\right)\delta^{(4)}(x-z)u_\mu u_\nu, 
\end{align}
and similarly, the scalar field equation~(\ref{field_eq}) changes to
\begin{align}
\Box\phi = -\gamma \mathcal{L}_{GB} f_{\phi} + \frac{m g_s}{\sqrt{-g}M_{\textrm{Pl}}}\int d\lambda \delta^{(4)}(x-z),
\end{align}
where we have used the normalization condition of the 4-velocity, $u^\mu u_\mu = -1$.

The equation of motion for the particle, obtained by varying the action~(\ref{eq:lagcp}) with respect to its worldline and taking the parameter $\lambda$ as the particle's proper time $\tau$, results in:
\begin{align}
u^\mu \nabla_\mu u^\nu & =- \frac{g_s}{M_{\textrm{Pl}}\left(1+\frac{g_s\phi}{M_{\textrm{Pl}}}\right)} (g^{\mu\nu} + u^\mu u^\nu) \nabla_\mu \phi.
\label{eq:geomp}
\end{align}
Notice that, in the limit $g_s=0$, the standard geodesic equation is recovered. However, for $g_s\neq0$, the right-hand side introduces a scalar-mediated force, proportional to the gradient of the scalar field. This force effectively acts as an additional ``attractive'' interaction for positive $g_s$, modifying the particle’s trajectory away from geodesic motion.

To compute the innermost stable circular orbit, we assume that the motion of the test particle has a negligible effect on the background geometry and scalar field. Under this assumption, the field equations remain those solved in Sec.~\ref{Sec.EGBss}, and the trajectory of the test particle is governed by Eq.~\eqref{eq:geomp}, which describes motion in the fixed background spacetime. 

We analyze circular orbits using the effective potential approach. Rather than solving the modified geodesic equations~\eqref{eq:geomp} directly, we make use of the normalization condition of the four-velocity, $u^\mu u_\mu = -1$, to compute the radial component of the four-velocity, $u_r$. This yields:
\begin{align}
u_r^2 = \frac{N^2 u_t^2 - r^2 u_\phi^2 - 1}{g^2}, \label{eq:unorm}
\end{align}
where, without loss of generality, we have restricted the motion to the equatorial plane by setting $\theta = \pi/2$.

From the action~(\ref{eq:lagcp}) and the stationary and axisymmetric ansatz~(\ref{Eq.StatAnz}), it follows that the temporal ($t$) and azimuthal ($\varphi$) coordinates are cyclic. Consequently, their associated conjugate momenta are conserved. Using these conserved quantities, we compute the temporal ($u^t$) and azimuthal ($u^\varphi$) components of the four-velocity:
\begin{subequations}\label{eq:cq}
\begin{align}
\mathcal{E} &:= -\frac{p_t}{m} = -\frac{1}{m} \frac{\partial L}{\partial u^t}
\quad \Rightarrow \quad
u^{t} = \frac{\mathcal{E}}{N^2} \left(1 + g_s \frac{\phi}{M_{\textrm{Pl}}} \right)^{-1}, \\
\ell &:= \frac{p_{\varphi}}{m} = \frac{1}{m} \frac{\partial L}{\partial u^\varphi}
\quad\quad \Rightarrow \quad
u^{\varphi} = \frac{\ell}{r^2} \left(1 + g_s \frac{\phi}{M_{\textrm{Pl}}} \right)^{-1},
\end{align}
\end{subequations}
where $L$ is the Lagrangian corresponding to the point-particle term in the action~(\ref{eq:lagcp}), and $\mathcal{E}$ and $\ell$ are the energy and angular momentum, respectively, both defined per unit rest mass.

Substituting the expressions for $u^t$ and $u^\varphi$ from Eq.~(\ref{eq:cq}) into the normalization condition~(\ref{eq:unorm}), we obtain the radial equation governing the motion of a test particle:
\begin{widetext}
\begin{align}
u_r^2 + V_{\text{Eff}}(r) = \mathcal{E}^2, \label{eq:req}
\end{align}
where
\begin{align}
V_{\text{Eff}}(r;\ell^2, \mathcal{E}^2, g_s) := \frac{1}{g^2}\left\{1 + \frac{\ell^2}{r^2}{ \left(1 + g_s \frac{\phi}{M_{\textrm{Pl}}} \right)^{-2}} + \mathcal{E}^2\left[g^2 - \frac{1}{N^2} {\left(1 + g_s \frac{\phi}{M_{\textrm{Pl}}} \right)^{-2}}\right]\right\},\label{eq:weff}
\end{align}
\end{widetext}
is the effective potential governing the radial motion of the particle.
Note that the last term disappears in the limit $\phi \to 0$ and $N \to 1/g$ (the Schwarzschild solution limit is recovered).

The existence of a stable circular orbit at a radius $r_0$ requires the following conditions:
\begin{align}\label{Eq.PotCond}
V_{\text{eff}}(r_0) = \mathcal{E}^2, \quad V_{\text{eff}}'(r_0) = 0, \quad V_{\text{eff}}''(r_0) > 0,
\end{align}
where the first condition ensures vanishing radial velocity ($u^r = 0$), the second corresponds to zero radial acceleration---necessary for the particle to remain in a circular orbit---and the third guarantees the stability of the orbit. If the last inequality is not satisfied, the orbit is unstable. (The marginal case $V_{\text{eff}}''(r_0) = 0$ requires a more detailed analysis, but this will not be necessary in the context of our study.)

Combining the two equality conditions from Eq.~(\ref{Eq.PotCond}), we obtain the following relation:
\begin{align}\label{Eq.Condit}
    \frac{N'(r_0)}{N(r_0)}=\frac{g_s\left(\frac{\ell^2}{g_s \left(1 + g_s \frac{\phi(r_0)}{M_{\textrm{Pl}}}\right) r_0^3}-\frac{\phi'(r_0)}{M_{\textrm{Pl}}}\right)}{\left(1 + g_s \frac{\phi(r_0)}{M_{\textrm{Pl}}} \right)\left(1+\frac{\ell^2}{r_0^2\left(1 + g_s \frac{\phi(r_0)}{M_{\textrm{Pl}}} \right)^2}\right)},
\end{align}
which implicitly determines the circular orbit radius $r_0$. The solutions are required to satisfy the non-degeneracy condition: $r_0^3 g(r_0)^2\left(1 + g_s \frac{\phi(r_0)}{M_{\textrm{Pl}}} \right)^2 N(r_0)\neq 0$, ensuring that the denominator remains finite and the circular orbit is well-defined. Using the numerical profiles corresponding to the Einstein–Gauss–Bonnet-scalar star solutions identified in Sec.~\ref{EGBSS}, one can numerically solve Eq.~\eqref{Eq.Condit} to determine the value of $r_0$. Note that in terms of the dimensionless variables~(\ref{Eq.Scaling}), we have $\bar{\ell}= \left(\lambda_{*} M_{\rm{Pl}}\right)^{-1} \ell$ and $\bar{g}_s=\lambda_{*} g_s$.

Figure~\ref{fig:lvsrmin} presents families of stable circular orbits as a function of the coupling parameter $g_s$ and angular momentum $\ell$, for two stellar configurations with mass $M_{99} = 37.81$. (Their respective mass profiles are shown in Fig.~\ref{fig:scalarMass}.) We consider a set of coupling values $g_s = 0, 0.02, 0.1, 0.3, 0.4, 0.5$, and employ Powell's dog-leg method (also known as Powell's hybrid method) to numerically identify the roots of the equality condition~(\ref{Eq.Condit}). The stability of the solutions is confirmed by applying the inequality condition Eq.~(\ref{Eq.PotCond}) to select only those roots that correspond to stable (circular) orbits. Left panel correspond to a typical boson stars (which is a solution of the Einstein-Klein-Gordon system), while the right panel shows the equivalent analysis but now considering a pure Einstein-scalar-Gauss-Bonnet star solutions: $\phi_{\infty}=0.01, a_1=2.0, \beta=10^3$. For comparison, we also show the locations of stable circular orbits around a Schwarzschild black hole, whose smaller radius is given by the innermost stable circular orbit at $r_{\text{ISCO}} = 6M/(8\pi)$.

To conclude, we offer a few general observations that can be drawn from Fig.~\ref{fig:lvsrmin}:
\begin{itemize}
\item[i.] Stable circular orbits exist for all values of $\ell$. This contrasts with the Schwarzschild case, where such orbits are only allowed if $\ell \geq \sqrt{3}M/(4\pi)$ and $r \geq r_{\text{ISCO}}$.
\item[ii.] When the particle is not coupled to the scalar field ($g_s = 0$), the radius of the stable circular orbit decreases with decreasing angular momentum. As $\ell \to 0$, the particle approaches $r \approx 0$. However, this limiting case does not satisfy the non-degeneracy condition discussed above, and thus the equality condition is not strictly applicable.
\item[iii.] For any $g_s \neq 0$, particles with zero angular momentum can remain stationary at a finite radius. This effect arises from the scalar field–particle coupling and has no analogue in the Schwarzschild case. However, it can be compared to the behavior observed in boson stars, as illustrated in the left panel of Fig.~\ref{fig:lvsrmin}.
\item[iv.] In contrast to Einstein–Gauss–Bonnet-scalar stars, for boson stars (i.e. within Einstein–Klein–Gordon gravity), a non-vanishing coupling $g_s \neq 0$ does not necessarily guarantee that a particle with zero angular momentum ($\ell = 0$) can remain stationary at a finite radius ($r \neq 0$). Instead, a minimum coupling strength is required for this behavior to emerge. For the boson star configuration shown in the left panel of Fig.~\ref{fig:lvsrmin}, this threshold occurs at approximately $g_s \gtrapprox 0.42$.
\end{itemize}
These observations highlight distinctive features of Einstein–Gauss–Bonnet scalar stars, particularly the existence of stable circular orbits at radii smaller than the ISCO of a Schwarzschild black hole, and the striking possibility that particles with non-zero scalar coupling and zero angular momentum can always remain stationary at finite radii. An effect enabled by the scalar–matter interaction and absent in standard general relativity.

%%%%%%%%%%%%%%%%%%%%%%%%%%%%
\section{Concluding remarks}\label{Sec.ConcRemark}
%%%%%%%%%%%%%%%%%%%%%%%%%%%%

In this paper, we investigate the existence and properties of compact objects within Einstein-scalar-Gauss-Bonnet gravity, with a non-minimally coupled massless real scalar field. We identify and numerically construct two main families of solutions: (i) \textbf{pure Einstein-scalar-Gauss-Bonnet stars}, which are supported solely by the scalar degree of freedom and characterized by a divergent scalar field at the origin while maintaining a globally regular geometry; and (ii) \textbf{hybrid Einstein-scalar-Gauss-Bonnet stars}, composed of both scalar and fermionic matter fields. These hybrid stars include branches with both regular and non-regular scalar field profiles at the origin, with the latter corresponding to a solution that has not been reported in the literature.

A key finding is that the non-regular scalar field solutions---despite their divergence at the origin---yield finite curvature invariants and asymptotically recover the Schwarzschild spacetime behavior. As can be concluded from Eq.~(\ref{Eq.ScCurvOrig}), these configurations do not possess analogs in standard general relativity, highlighting the unique role played by the Gauss-Bonnet coupling.

From our phenomenological study, we find that both pure scalar and hybrid stars exhibit a rich behavior. Pure scalar stars can reach compactness values that exceed those of standard boson stars and even the Buchdahl limit for fluid stars, while remaining below the Schwarzschild black hole threshold. Their high compactness suggests potential as gravitational wave super-emitters. Hybrid stars, on the other hand, can mimic neutron star-like masses even when modeled with simple polytropic equations of state. For the regular scalar branch, they smoothly connect to general relativity in the limit $\beta \to 0$. In contrast, the non-regular scalar branch does not have such a limit and may offer alternative astrophysical signatures.

We also examined the motion of test particles non-minimally coupled to the scalar field and showed that such couplings allow for qualitatively new behaviors, such as stable circular orbits within the Schwarzschild's ISCO and static configurations at finite radii for particles with zero angular momentum.

Altogether, our findings illustrate that Einstein–Gauss–Bonnet gravity supports a rich landscape of compact objects with distinctive phenomenological properties, offering new opportunities to test gravitational theories in the strong-field regime. This opens multiple avenues for future research, including linear stability, the inclusion of more realistic equations of state, and scalarization processes in the presence of matter, which we will explore in future work.

%%%%%%%%%%%%%%%%%%%%%%%%%%%%
\begin{acknowledgments}
A.A.R. acknowledges funding from a postdoctoral fellowship from ``Estancias Posdoctorales por México para la Formación y Consolidación de las y los Investigadores por México''. J.C. by CONACyT/DCF/320821.
\end{acknowledgments}

%%%%%%%%%%%
\appendix
%%%%%%%%%%%

\section{Complementary equations}\label{Ap.Compl.Eq}

In Sec.~\ref{Sec.EGBg}, the Einstein-Gauss-Bonnet field equations are presented. Below, we provide the explicit expressions for the tensors $\alpha_{\mu\nu}$, and $\beta_{\mu\nu}$ used in Eq.~(\ref{Eq.GBContrib}):
\begin{widetext}
\begin{eqnarray}
\alpha_{\mu\nu} &=& -\frac{R}{2}\nabla_\mu\nabla_\nu\phi-G_{\mu\nu}\Box\phi+{R_{\nu}}^{\gamma}\nabla_{\gamma}\nabla_{\mu}\phi+{R_{\mu}}^{\gamma}\nabla_{\gamma}\nabla_{\nu}\phi+(R_{\mu\gamma\nu\beta}-g_{\mu\nu}R_{\gamma\beta})\nabla^{\beta}\nabla^{\gamma}\phi \,,\label{Eq.Ap.Covar1}\\
\beta_{\mu\nu} &=& -\frac{R}{2}\nabla_\mu\phi\nabla_\nu\phi+(R_{\nu\gamma}\nabla_{\mu}\phi+R_{\mu\gamma}\nabla_{\nu}\phi-G_{\mu\nu}\nabla_{\gamma}\phi)\nabla^{\gamma}\phi + (R_{\mu\gamma\nu\beta}-g_{\mu\nu}R_{\gamma\beta})\nabla^{\gamma}\phi\nabla^{\beta}\phi \,.\label{Eq.Ap.Covar3}
\end{eqnarray}
\end{widetext}
As usual, $R$, $R_{\mu\nu}$, $R_{\mu\nu\gamma\beta}$, and $G_{\mu\nu} := R_{\mu\nu} - \frac{1}{2} g_{\mu\nu} R$ denote the Ricci scalar, Ricci tensor, Riemann tensor, and Einstein tensor, respectively, all defined with respect to the spacetime metric $g_{\mu\nu}$. The D'Alembert operator is defined as $\Box := \nabla_{\mu} \nabla^{\mu}$.

%%%%%%%%%%%%%%%%
\section{Hybrid Einstein-scalar-Gauss-Bonnet star solutions with a radial scalar field that has a minimum at the origin}\label{Ap.Demostrat}
%%%%%%%%%%%%%%%%

In Sec.~\ref{Sec.RegScalFieldatOrig}, we present theoretical hybrid Einstein-scalar-Gauss-Bonnet star solutions featuring a regular scalar field at the origin. A distinctive feature of these solutions is a radial scalar field that has a minimum at the origin and asymptotically approaches a nonzero constant value. Below, we provide an intuitive demonstration of this behavior near the origin when $\beta>0$.

Using the system~(\ref{Eq.OrigSecDeriva}) one can derive the following relation:
\begin{align}
\frac{\phi_{0}''}{M_{\rm{Pl}}^2} = \frac{1 \pm \sqrt{1 - 128\left(\frac{\gamma f_{\phi_0} g_{0}''}{M_{\rm{Pl}}}\right)^2\left(1 + \frac{p_0}{M_{\rm{Pl}}^2 g_0''}\right)}}{16 \gamma f_{\phi_0}}.
\end{align}
Considering scalar coupling functions such that $f_{\phi_0} > 0$, as in our case (see Eq.~(\ref{Eq.FuncPhi})), it is straightforward to show that the scalar field always exhibits a local minimum at $r = 0$ (i.e., $\phi_0' = 0$, $\phi_0'' > 0$) when $g_0'' > 0$. (Recall that regularity at the origin requires $\phi_0' = 0$.) In contrast, when $f_{\phi_0} < 0$, the scalar field always exhibits a local maximum at the origin.

Regularity at the origin requires that $g_0 = 1$ and $g_0' = 0$ (see Eqs.~(\ref{Eq.RicciCond},~\ref{Eq.BoundCond})). Since we recover Schwarzschild behavior asymptotically, $\lim_{r \to \infty} g = 1$, any non-constant solution must exhibit a $g$-metric profile that either increases or decreases near the origin. Our numerical results show that, when the scalar field is regular (at the origin), the metric function $g(r)$ upward near the origin (i.e., $g_0'' > 0$; see Fig.~\ref{fig:MIllExamp}). In contrast, when the scalar field is non-regular, $g(r)$ decreases near the origin (see Fig.~\ref{fig:MIllExamp2}). This behavior explains why, in the regular case, the scalar field exhibits a local minimum at the origin.

%%%%%%%%%%%%%%%%
%\section*{References}
% \bibliographystyle{unsrt}
%%%%%%%%%%%%%%%%
\bibliography{ref.bib}

\end{document}